%% file: LatexDM2TNet.tex
\documentclass[10pt,journal]{IEEEtran}
\usepackage{amsmath,amsfonts}
\usepackage{algorithm}
\usepackage{array}
\usepackage{textcomp}
\usepackage{stfloats}
\usepackage{url}
\usepackage{verbatim}
\usepackage{graphicx}
\usepackage{cite}
\usepackage{algorithm} 
\usepackage{algpseudocode}

\usepackage{caption}
\usepackage{subcaption}
\usepackage{url}  
\usepackage{enumitem}
\usepackage{enumerate}
\usepackage{color}
\usepackage{multirow}
\usepackage{amssymb}
\usepackage{booktabs}
\usepackage{algpseudocode}
\usepackage{bbold}
\usepackage{multirow}
\usepackage{bm}

\definecolor{dgreen}{rgb}{0.412,0.741,0.271}
\definecolor{dblue}{rgb}{0.220,0.325,0.639}
\definecolor{dred}{rgb}{0.933,0.122,0.137}

\newcommand{\wjc}[1]{{\textcolor{black}{#1}}}
\newcommand{\wjcfinal}[1]{{\textcolor{black}{#1}}}

\begin{document}

\title{Dual Multi-scale Mean Teacher Network for Semi-supervised Infection Segmentation in Chest CT Volume for COVID-19}

\author{
        Liansheng~Wang,
        Jiacheng~Wang,
        Lei~Zhu,
        Huazhu~Fu,
        Ping~Li,
        Gary~Cheng,
        Zhipeng~Feng,
        Shuo~Li,
        and~Pheng-Ann~Heng
       \IEEEcompsocitemizethanks{
       \IEEEcompsocthanksitem L. Wang and J. Wang are
       with Department of Computer Science at School of Informatics, Xiamen University.
       \IEEEcompsocthanksitem L. Zhu is with The Hong Kong University of Science and Technology (Guangzhou), Nansha, Guangzhou, 511400, Guangdong, China
       and The Hong Kong University of Science and Technology, Hong Kong SAR, China  (Email: leizhu@ust.hk)
       \IEEEcompsocthanksitem P.-A. Heng. is with  Department of Computer Science and Engineering, The Chinese University of Hong Kong.
       \IEEEcompsocthanksitem H.~Fu is with the Institute of High Performance Computing (IHPC), Agency for Science, Technology and Research (A*STAR), Singapore 138632. (E-mail: hzfu@ieee.org)
       \IEEEcompsocthanksitem P. Li is with the Department of Computing, The Hong Kong Polytechnic University, Kowloon, Hong Kong (Email: p.li@polyu.edu.hk).
      \IEEEcompsocthanksitem G. Cheng is with Department of Mathematics and Information Technology, The Education University of Hong Kong.
      \IEEEcompsocthanksitem Z. Feng is with Zhongshan Hospital, Xiamen University.
       \IEEEcompsocthanksitem S. Li is with Western University.
       \IEEEcompsocthanksitem Lei Zhu is the corresponding author of this work (leizhu@ust.hk).
        }
}

\markboth{IEEE TRANSACTIONS ON CYBERNETICS}%
{Shell \MakeLowercase{\textit{et al.}}: Bare Demo of IEEEtran.cls for Computer Society Journals}


\maketitle

\input{abstract}
\input{introduction}
\input{related_work}
\input{methods}
\input{experiments}
\input{conclusion}

\section*{Acknowledgments}
The work is supported by the National Natural Science Foundation of China (Project No. 61902275), and the AI Singapore Tech Challenge (Open-Theme) Funding (AISG2-TC-2021-003).

\bibliographystyle{IEEEtran}
\bibliography{LatexDM2TNet}
\vfill
\end{document}

%% file: abstract.tex
\begin{abstract}
Automated detecting lung infections from computed tomography (CT) data plays an important role for combating COVID-19. However, there are still some challenges for developing AI system. 1) Most current COVID-19 infection segmentation methods mainly relied on 2D CT images, which lack 3D sequential constraint. 2) Existing 3D CT segmentation methods focus on single-scale representations, which do not achieve the multiple level receptive field sizes on 3D volume. 3) The emergent breaking out of COVID-19 makes it hard to annotate sufficient CT volumes for training deep model.
To address these issues, we first build a multiple dimensional-attention convolutional neural network (MDA-CNN) to aggregate multi-scale information along different dimension of input feature maps and impose supervision on multiple predictions from different CNN layers. Second, we assign this MDA-CNN as a basic network into a novel dual multi-scale mean teacher network (DM${^2}$T-Net)  for semi-supervised COVID-19 lung infection segmentation on CT volumes by leveraging unlabeled data and exploring the multi-scale information.
Our DM${^2}$T-Net encourages multiple predictions at different CNN layers from the student and teacher networks to be consistent for computing a multi-scale consistency loss on unlabeled data, which is then added to the supervised loss on the labeled data from multiple predictions of MDA-CNN.
\wjc{Third, we collect two COVID-19 segmentation datasets to evaluate our method. The experimental results show that our network consistently outperforms the compared state-of-the-art methods.}
\end{abstract}

%% file: introduction.tex
\section{Introduction}
\label{sec::introduction}

As an ongoing pandemic, novel coronavirus 2019 (COVID-19) has infected about 6,898,613 cases and incurred 399,832 deaths in the world by 7th June, 2020.
Reverse Transcription-Polymerase Chain Reaction (RT-PCR) test is considered as the gold standard of screening COVID-19.
However, RT-PCR testing is time-consuming and requires repeated
testing for accurate confirmation of a COVID-19 case due to its low sensitivity, thereby resulting in the ineffectiveness of timely confirming CVOID-19 patients.
By working as a complement to RT-PCR,  easily accessible imaging equipments (e.g., chest X-ray and computed tomography (CT)), have provided huge assistance to clinicians in both current diagnosis and follow-up assessment of disease evolution~\cite{shi2020review,rubin2020role,fan2020inf}.
Further, quantitative CT information (e.g., lung burden, percentage of high opacity, and lung severity score) are used widely to monitor disease progression and understand the course of COVID-19~\cite{chassagnon2020ai,chaganti2020quantification,zheng2020predicting}.
In clinical practice, CT screening is usually more preferred since typical infection signs can be observed from CT data, covering ground-glass opacity (GGO) in the early stage to pulmonary consolidation in the late stage.

Moreover, the qualitative evaluation of infection and longitudinal changes in CT images could thus provide useful and important information in combating COVID-19.

Accurate segmentation of the COVID-19 infected region plays a crucial role in achieving a reliable quantification of infection in chest CT images. However, the manual delineation of lung infections is tedious, labor-consuming, and expensive for radiologists.
Also, annotating the lung infections in CT is a challenging task due to highly variant textures, sizes and positions of infected regions, as well as low contrast and blurred GGO boundaries~\cite{chen2020residual}.
Recently, convolutional neural networks (CNNs) have been developed to automatically segment the lung infections from CT images~\cite{zhou2020automatic,yan2020covid,qiu2021miniseg,chen2020residual,xie2020relational,fan2020inf,Kang2020TMI}. 
\wjc{Such automatic segmentation tools contribute to the quantification of lung infections and can eliminate the possibility of subjective impact. Moreover, for saving medical resources and accelerating daily diagnosis of the overburdened hospitals under the COVID-19's large-scale outbreak, building AI system will be greatly helpful.}
However, there are still some challenges for developing AI system.
1) Most current COVID-19 infection segmentation methods mainly relied on 2D CT images, resulting in a lower  segmentation accuracy due to lack inter-relations among different 2D images of raw clinical CT volume.
Moreover, infected regions of a 3D volume only exist in a few 2D images and using 2D images to train a segmentation network tends to produce many false positives in these images without any lung infection.
\wjc{By contrast, the 3D context makes a meaningful difference in exploring inter-relation between continuous slices and determining the infection regions from features of more dimensions so as to increase the segmentation accuracy.}
2) The 3D volume segmentation methods~\cite{ma2020towards,ronneberger2015u} only exploited the single scale of the input 3D CT volume, which do not consider the multiple level receptive field sizes on 3D volume.
3) It is difficult to collect sufficient high-quality labeled 3D CT data within a short time for training a deep model, which limits the developing of deep 3D segmentation models.


To address above issues, we propose a dual multi-scale mean teacher network (DM$^2$T-Net) for boosting the 3D COVID-19 lung infection segmentation performance. As shown in Fig.~\ref{fig:arc}, our  DM$^2$T-Net utilizes two kinds of multi-scale structures. One is the multiple dimensional-attention for learning a hierarchical representations of 3D volume, while another is multi-scale consistency loss for constraining the semi-supervised learning.
Specifically, a novel multiple dimensional-attention convolutional neural network (MDA-CNN) is designed as the basic network for both teacher and student networks of DM$^2$T-Net.
The proposed MDA-CNN attentively integrates multiple scales of the input 3D volumes along different dimensions for segmenting lung infections, simultaneously, produces multiple side-outputs at different CNN layers.
Two kinds of loss are employed to constrain our DM$^2$T-Net. One is multi-scale supervised loss in the student network for labeled data to integrate the deeply-supervised side-outputs. The second is multi-scale consistency loss for the unlabeled data to encourage multiple predictions at different CNN layers from the student and teacher networks to be consistent.
%
%
Overall, the main contributions are summarized as:
\begin{itemize}
\item First, we develop a multiple dimensional-attention convolutional neural network (MDA-CNN) for 3D lung infection segmentation,
which attentively aggregated CNN features extracted from multiple dimensional-scale information of the input 3D data and generates multiple predictions at different CNN layers.
\item Second, we propose a dual multi-scale mean teacher network (DM${^2}$T-Net) for leveraging the unlabeled data.
A multi-scale consistency loss is devised on the side-output predictions to encourage the predictions consistent on both inter-model and intra-model.
As a semi-supervised learning model, our framework has the potential to be used for other 3D segmentation tasks.
\item Moreover, we collect two COVID-19 segmentation datasets to evaluate our method.
The experiments show that our proposed network outperforms state-of-the-art methods on both supervised and semi-supervised manners. \wjcfinal{We have released the code, trained models, and collected unlabeled 3D CT data at \url{https://github.com/jcwang123/DM2TNet}.}
\end{itemize}

%% file: related_work.tex
\section{Related Work}
\label{sec:related_work}
\begin{figure*} [t]
	\centering
	\includegraphics[width=0.9\linewidth]{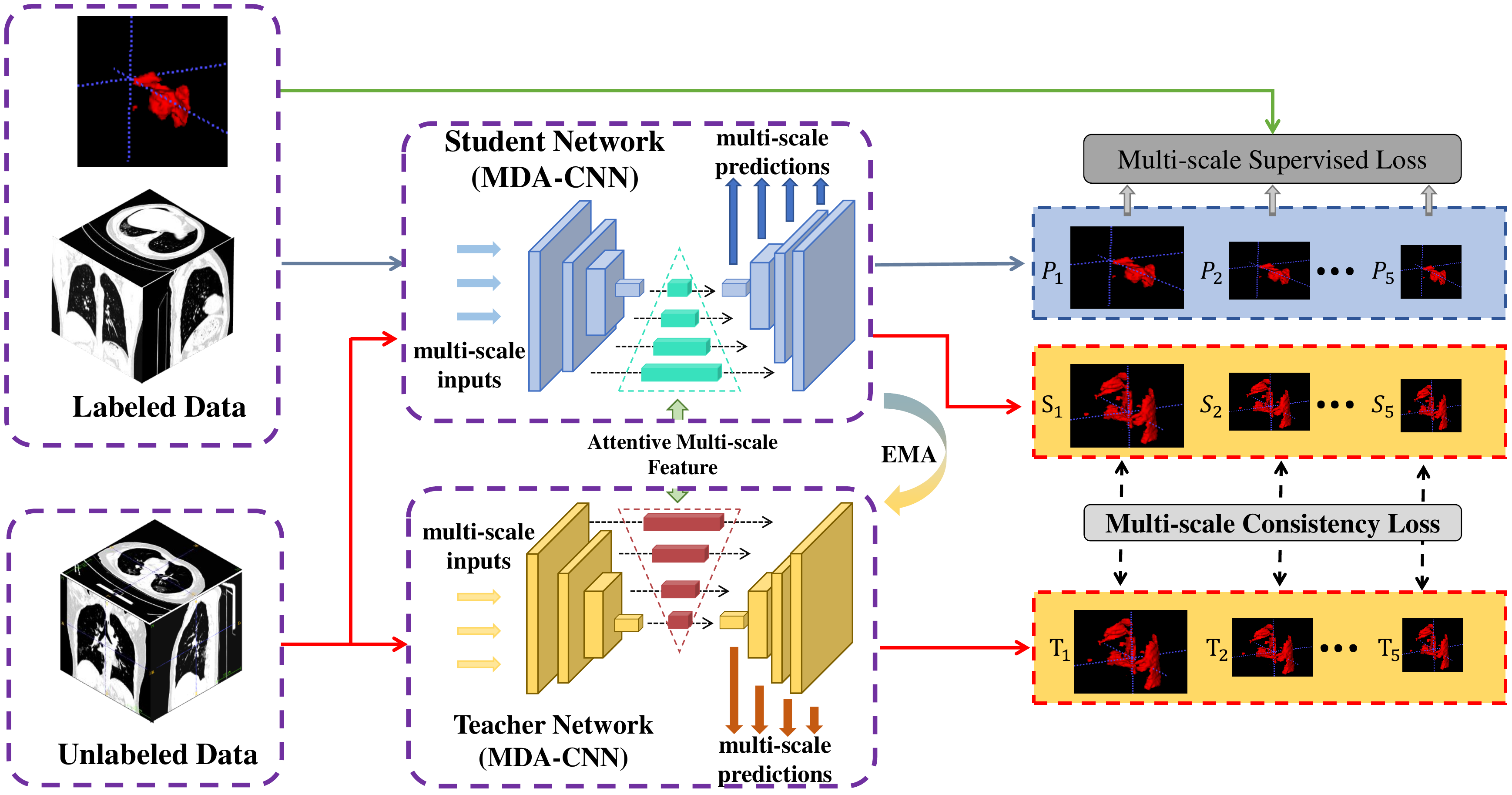}
	\caption{The schematic illustration of the developed dual multi-scale mean teacher network (DM$^2$T-Net). We first develop a multiple dimensional-attention convolutional neural network (MDA-CNN; see Fig.~\ref{fig:arc-MDCNN}) to detect infected regions. The MDA-CNN produces five segmentation results from different CNN layers. After that, we compute a multi-scale supervised loss on the five segmentation results (i.e., $P_1$ to $P_5$) of unlabeled data. For unlabeled data, we compute a multi-scale supervised loss on two pairs of five results from the student network (i.e., $S_1$ to $S_5$) and the teacher network (i.e., $T_1$ to $T_5$). Finally, we combine the supervised loss and consistency loss to train our 3D COVID-19 infection segmentation network.}
	\label{fig:arc}
\end{figure*}

This section reviews three kinds of works that are most
related to our method, including segmentation in chest CT, semi-supervised learning, and data-driven methods for COVID-19.

\subsection{Segmentation in Chest CT}
Segmenting organs and tumors from chest CT images provides crucial information for clinicians to diagnose and quantify lung diseases~\cite{gordaliza2018unsupervised,munoz2012quantification}.
Early data-driven algorithms converted the lung segmentation task into voxel classification and then applied different classification models with manually-designed features for  segmenting target regions.
Wu et al.~designed a set of texture and shape features to represent voxels and then trained conditional random field (CRF) model to classify voxels for lung segmentation~\cite{wu2010stratified}.
Keshani et al.~segmented lung nodules by the support vector machine (SVM) classifier with 2D stochastic and 3D anatomical features~\cite{keshani2013lung}.
However, relying on these hand-crafted features is difficult to segment nodules due to similar appearances of nodules and background details.
Motivated by outstanding performance of CNNs in medical image analysis~\cite{LeiLei7815379,WuWu7875138,ChenChen7890445,ShengSheng8362709,8917648,fu2019angle,liu2019weakly}, deep learning based methods have been introduced to learn discriminative representation for lung nodule detection from CT images.
Wang et al.~formulated a central focused convolutional neural
network to capture both 2D and 3D lung nodule features for identifying lung nodules from heterogeneous CT images~\cite{wang2017central}.
Jin et al.~developed a conditional GAN model to generate CT-realistic high-quality 3D lung nodules~\cite{jin2018ct} and utilized these synthesized data to enhance the pathological lung segmentation model~\cite{harrison2017progressive}.
Jiang et al.~simultaneously leveraged features across multiple image resolution and CNN feature levels via residual connections to identify the lung tumors~\cite{jiang2018multiple}.

\subsection{Semi-supervised Learning}
Annotations in large-scale medical data are tedious, time-consuming and difficult to obtain.
More and more researchers have shifted their attentions from  supervised fashion to the semi-supervised learning, which improves the model performance by combining limited labeled data and sufficient unlabeled data~\cite{zhu2005semi}.
From a high-level view, these semi-supervised learning methods devised an objective function, which consists of supervised loss on labeled data and unsupervised learning on unlabeled data or both the labeled data and unlabeled data.
Lee et al.~picked up the class which has the maximum predicted probability and used this class as the pseudo-Labels for unlabeled data~\cite{lee2013pseudo}.
Bai et al.~alternately update the network parameters and the segmentation on unlabeled data in a semi-supervised learning framework~\cite{bai2017semi}.
Based on an adversarial learning based semi-supervised fashion~\cite{dong2018unsupervised,nie2018asdnet},  Zhang et al.~encouraged the segmentation of unlabeled images to be similar to those of the labeled ones in a deep adversarial network~\cite{zhang2017deep}.
Yu et al.~estimated an uncertainty information as a guidance to eliminate unreliable predictions and maintain only the reliable ones (low
uncertainty) when devising the consistency loss of student and teacher network predictions for labeled and unlabeled data~\cite{yu2019uncertainty}.

\subsection{AI Techniques for COVID-19}

Artificial intelligence (AI) methods, especially deep learning techniques, have been employed widely in medical imaging applications against COVID-19~\cite{shi2020review,CVOID_review2}.
Tang et al.~calculated quantitative features from chest CT images and then passed these features into to train a random forest model for COVID-19 severity assessment~\cite{tang2020severity}.
Wang et al.~modified an inception
transfer-learning model for the identification of viral pneumonia images~\cite{wang2020deep}.
Chen et al.~first collected 46,096
image slices from 106 admitted COVID-19 patients,
and then trained U-Net++~\cite{zhou2018unet++} to extract valid areas and detect suspicious lesions in CT images~\cite{chen2020deep}.
Song et al.~formulated a detail relation extraction neural network (DRE-Net) to extract top-K details and obtain the image-level predictions for patient-level diagnoses~\cite{song2020deep}.
Xie et al.~presented a relational two-stage U-Net to segment pulmonary lobes in CT images by introducing a non-local
neural network module to model the global structured
relationships~\cite{xie2020relational}.
Chen et al.~exploited both the residual network
and attention mechanism to improve the efficacy of the U-Net for the lung CT image segmentation~\cite{chen2020residual}.
Wu et al.~created a COVID-19 dataset with $3,855$ labeled CT images and performed a joint explainable Classification and accurate lesion segmentation~\cite{wu2020jcs}.
Qiu et al.~presented a lightweight deep learning model for efficient COVID-19 image segmentation~\cite{qiu2021miniseg}.
Observing that the boundary of the infected region can be enhanced by adjusting the global intensity, Yan et al.~introduced a deep CNN with  feature variation block, which adaptively adjusted the global properties of the features for segmenting COVID-19 infection in 2D images~\cite{qiu2021miniseg}.
Zhou et al.~incorporated a spatial and channel attentions to U-Net model for capturing richer contextual relationships for COVID-19 CT segmentation~\cite{zhou2020automatic}.
He et al.~first used a set of 2D image patches to represent each CT image and then developed a multi-task multi-instance to assess the severity of COVID-19 patients and segment the lung lobe simultaneously~\cite{he2020synergistic}.
Fan et al.~devised an Inf-Net to utilize an implicit reverse attention and explicit edge-attention for the identification of infected regions and then further enhance the segmentation performance by embedding the Inf-Net into a semi-supervised learning strategy~\cite{fan2020inf}.
\wjcfinal{Ding et al.~ proposed the MT-nCov-Net to formulate 2D lesion segmentation as a multitask shape regression problem that enables the feature fusion between various tasks~\cite{ding2021mt}. 
Pang et al.~proposed a novel group equivariant segmentation framework by encoding the 2D inherent symmetries, i.e., rotations and reflections, for learning more precise representations~\cite{pang2022beyond}.}
Methods above almost relied on 2D image images to train CNNs for the classification or segmentation.
%
Ma et al.~created a COVID-19 3D CT dataset with 20 cases and explored the U-Net~\cite{cciccek20163d} for 3D Lung and Infection Segmentation~\cite{ma2020towards}.

%% file: methods.tex
\section{Proposed Method}
\label{sec:method}

\begin{figure*} [!t]
	\centering
	\includegraphics[width=1.0\linewidth]{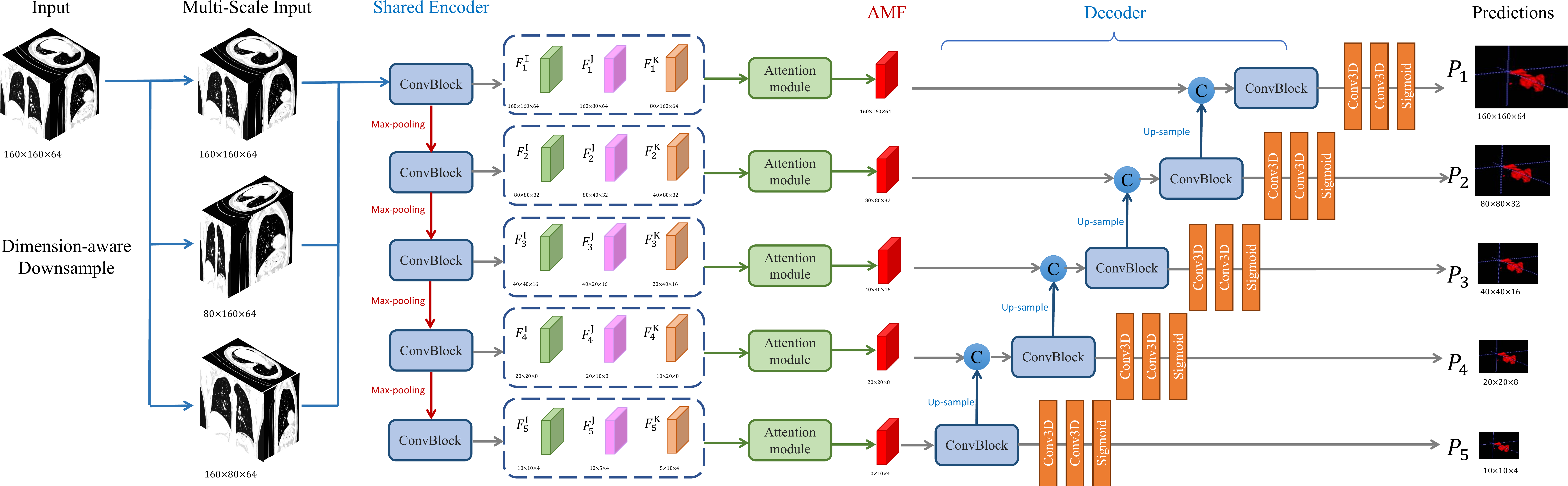}
	\caption{The schematic illustration of our multiple dimensional-attention convolutional neural network (MDA-CNN).
	Given an input 3D CT volume, MDA-CNN first down-samples it along two spatial dimensions and then passes the three 3D volumes to obtain a set of CNN features ($F_i^j$) with different spatial resolutions.
	Then, the attention modules are introduced after each CNN layer to attentively aggregate features ($AMF_i$) from three volumes and these aggregated features are then adjacently merged.
	Finally, multiple side-outputs ($P_i$) are produced form different decoder layers.
	Note that we empirically take $P_1$ as the output segmentation result of MDA-CNN.
	}
	\label{fig:arc-MDCNN}
\end{figure*}

Fig.~\ref{fig:arc} shows the schematic illustration of the proposed DM$^2$T-Net which integrates dual multi-scale details of labeled data and unlabeled data for COVID-19 lung infection segmentation. In our DM$^2$T-Net, a multiple dimensional-attention convolutional neural network (MDA-CNN) is developed to attentively aggregate CNN features extracted from multiple dimensional-scale information and produce multiple predictions from different CNN layers.
This MDA-CNN is then integrated into DM$^2$T-Net as the basic network for both the student and the teacher networks.
In the training stage, the labeled data is fed into the student network, and a multi-scale supervised loss is calculated to constrain the consistency of multiple side-outputs on intra-model.
Then, the unlabeled data is inputted into the both student and teacher networks, respectively. Meanwhile, a multi-scale consistency loss is devised on the two groups of side-outputs to encourage the predictions consistent on inter-model.

\subsection{Multiple Dimensional-attention CNN}
\label{subsec:MDA-CNN}

Although achieving remarkable results, existing 3D segmentation networks produce unsatisfactory results when detecting 3D lung infected regions, since only single scale information of input volume is considered.
To address this issue, we argue that exploring multi-scale information is helpful to boost lung infection segmentation in 3D CT volume.
In this paper, we propose a multiple dimensional-attention convolutional neural network (MDA-CNN) to model and fuse the complementary of multiple dimensional-scale details within a single network, as shown in Fig.~\ref{fig:arc-MDCNN}.
Given a 3D lung CT volume ($\mathcal{I}$), we first generate another two auxiliary volumes (denoted as $\mathcal{J}$ and $\mathcal{K}$) by down-sampling $\mathcal{I}$ along the two spatial dimensions.
Then, we pass the three volumes into several convolutional blocks and obtain multiple feature maps with five different spatial resolutions for each volume ($\mathcal{I}$, $\mathcal{J}$ and $\mathcal{K}$).
We use ${F_{i}}^{\mathcal{I}}$, ${F_{i}}^{\mathcal{J}}$, and ${F_{i}}^{\mathcal{K}}$ to denote the three features at the $i$-th CNN layer; see Fig.~\ref{fig:arc-MDCNN}.
After that, we develop an attention module in each CNN layer to learn attention maps for aggregating the features with multiple dimension-aware scale information.
By doing so, the dimension-aware multi-scale representations of the input volume are well modeled and fused together for segmenting COVID-19 infected lung areas.


\begin{figure} [t]
	\centering
	\includegraphics[width=1.0\linewidth]{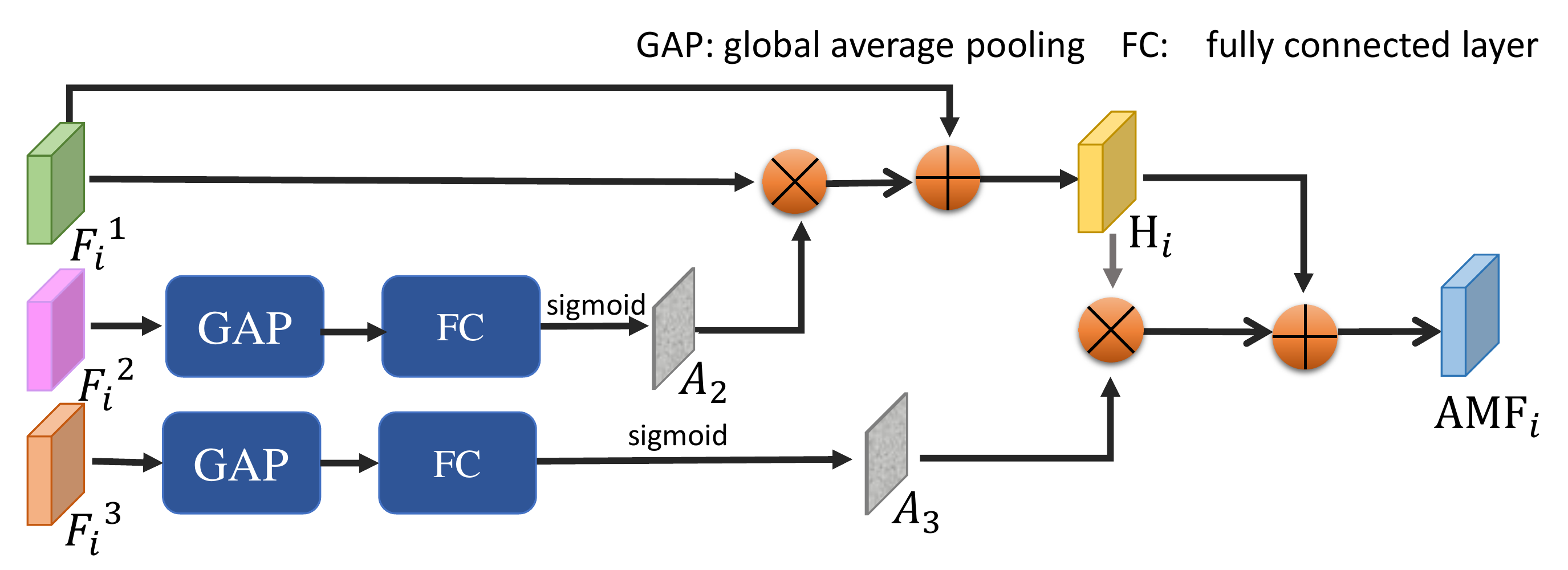}
	\caption{The schematic illustration of the developed attention module of Fig.~\ref{fig:arc-MDCNN}}
	\label{fig:attention}
\end{figure}

Fig.~\ref{fig:attention} shows the schematic illustration of the attention module at the $i$-th CNN layer. It takes three feature maps (${F_{i}}^{\mathcal{I}}$, ${F_{i}}^{\mathcal{J}}$, and ${F_{i}}^{\mathcal{K}}$) from $\mathcal{I}$, $\mathcal{J}$, and $\mathcal{K}$ at the $i$-th CNN layer and outputs a new feature map (${AMF}_i$) to attentively aggregate the input features.
The attention module starts by using one average pooling operation (GAP) on ${F_{i}}^{\mathcal{J}}$, one fully connected (FC) layer, and one Sigmoid activation function to obtain an attention map ($A_2$).
Then, we multiply the attention map $A_2$ with ${F_{i}}^{\mathcal{I}}$, and the resultant features are then element-wisely added with ${F_{i}}^{\mathcal{I}}$ to obtain a refinement of ${F_{i}}^{\mathcal{I}}$, which is denoted as $H_{i}$.
Similarly, we further employ an average pooling operation on ${F_{i}}^{\mathcal{K}}$ to obtain another attention map ($A_3$), which is then multiplied with $H_{i}$.
We then element-wisely add the resultant feature map to $H_{i}$ to obtain attentive multi-scale features (denoted as ${AMF}_i$), which is taken as the output of the developed attention module.
After the encoder layers and attention modules,
we obtain five attentive feature maps (denoted as ${AMF}_i$) with different scales $i=1,...,5$, which are then adjacently merged for decoding them.
Then, we use three $3\times3$ CNN layers, one $1$$\times$$1$ CNN layer, and one Sigmoid activation layer on each decoder output to produce five segmentation predictions, and then apply the deep supervision mechanism~\cite{xie2015holistically} to impose the supervision on each segmentation result.

\subsection{Multi-scale Supervised Loss on Labeled data}
\label{subsec:supervised loss}
Given a labeled data, we can have a pair of input 3D CT data and the corresponding annotated lung infection mask.
It is natural that we take the annotated infection mask as the ground truth of the COVID-19 infected region segmentation.
As shown in Fig.~\ref{fig:arc-MDCNN}, the proposed MDA-CNN predicts multiple segmentation results at different CNN layers and these results have different spatial resolutions; see $P_1$ to $P_5$ of the last column in Fig.~\ref{fig:arc-MDCNN}.
Hence, we down-sample the original segmentation mask into the same resolution of the prediction result at each CNN layer as its ground truth.
After obtaining the ground truths for the segmentation result at each CNN layer, we devise a  multi-scale supervised loss (denoted as $\mathcal{L}^{s}$) for a labeled image ($x_n$) by adding the supervised losses of all the CNN layers:
\begin{equation}\label{Eq:total_supervised_loss}
    \mathcal{L}^s(x_n)= \sum_{k=1}^5 \Phi_{dice} \left( P_k, G_k \right) \ ,
\end{equation}
where $P_k$ denotes the predicted the lung infection detection map at $k$-th CNN layer.
$G_k$ is the down-sampled ground truth at $k$-th CNN layer and it has the same resolution of $P_k$.
Here, we empirically use the dice loss to compute the difference between $P_k$ and $G_k$.

\subsection{Multi-scale Consistency Loss on Unlabeled Data} \label{subsec:consistency-loss}

The unlabeled 3D CT data was fed into the student network and teacher network to obtain their infection segmentation results, which are the five predictions at different CNN layers.
We then devise a a multi-scale consistency loss ($\mathcal{L}^c$) to enforce the five predictions from the student network and teacher network to be consistent.
The definition of $\mathcal{L}^c$ is given by:
\begin{equation}\label{Eq:total_consistency_loss}
    \mathcal{L}^c(y_m) = \sum_{k=1}^5 \Phi_{MSE} \left( S_{k}, T_{k} \right) \ \,
\end{equation}
where $y_m$ is the input unlabeled image. $S_{k}$ and $T_{k}$ are the segmentation results of the student network and the teacher network at $k$-th CNN layer. Here we empirically use the mean square error (MSE) loss to compute the difference of $S_{k}$ and $T_{k}$.

\wjc{Like the original mean teacher framework~\cite{tarvainen2017mean}, the multi-scale consistency loss on unlabeled data is designed to improve the segmentation performance. The reason why the consistency loss on unlabeled data can increase the segmentation accuracy is summarized as: our semi-supervised network first generates an initial pseudo label for unlabeled data by training the segmentation network on labeled data, and then computes the consistency loss on the predictions of a teacher network and a student network to progressively refine the pseudo labels for unlabeled data. By doing so, we can generate more reliable predictions on unlabeled data, and these reliable unlabeled data are combined with the labeled data into the training process, thereby boosting the overall segmentation performance.}

\subsection{Our Loss Function}
\label{subsec:our-network}

The total loss ($\mathcal{L}_{total}$) of our network is computed as:
\vspace{-2mm}
\begin{equation}\label{Equ:total_loss}
 \mathcal{L}_{total} = \sum_{n=1}^{N} \mathcal{L}^s(x_n) + \lambda \sum_{m=1}^{M} \mathcal{L}^c(y_m) \ ,
\end{equation}
where $N$ is the number of labeled CT scans of the training set.
$M$ is the number of unlabeled CT scans of our training set.
$\mathcal{L}^s(x_n)$ denotes the multi-scale supervised loss (Eq.~\eqref{Eq:total_supervised_loss}) for the $n$-th labeled image ($x_n$) while $\mathcal{L}^c(y_m)$ is the multi-scale consistency loss (Eq.~\eqref{Eq:total_consistency_loss}) for the $m$-th unlabeled image ($y_m$).
The weight $\lambda$ balances the multi-scale supervised loss of labeled data and the multi-scale consistency loss of unlabeled data.
As suggested in~\cite{yu2019uncertainty,tarvainen2017mean}, we compute $\lambda$ via a time-dependent Gaussian warming up function: $\lambda(i) = \lambda_{max} e^{(-5{(1-i/i_{max})}^2)}$, where $t$ denotes the current training iteration and $i_{max}$ is the maximum training iteration.
We empirically set $\lambda_{max}$$=$$5$ in our experiments.

\wjc{
Note that averaging model weights over CNN training steps tends to produce a more accurate model than using the final weights directly. Based on this, the mean teacher~\cite{tarvainen2017mean} computes the weights of a network (called teacher network) as an exponential moving average (EMA) weights of a network (called student network) to generate a better target model, which the student network learns from. Hence, one network learns from another network during the training process of the mean teacher framework, and the former network is named the student network, and the latter network is named the teacher network.
}
In the  $t$ training iteration, the teacher network parameters $\Omega'_t$ is computed by:
\begin{equation}\label{Equ:theta_update}
  \Omega'_t = \beta \Omega'_{t-1}+ (1-\beta)\Omega_t,
\end{equation}
where $\Omega_t$ is the student network parameter at the $t$ training iteration. We set EMA decay $\beta=0.99$ as same as in~\cite{yu2019uncertainty,tarvainen2017mean}.

\input{table-comparison-part1}

\input{table-comparison-part2}


%% file: table-comparison-part1.tex
\begin{table*}[t]
	\caption{\wjc{The results (mean $\pm$ variance) of different segmentation methods on \textbf{COVID-19-P20}}. We use the bold fonts to highlight the best performance.}
	\label{table:state-of-the-art}
	\vspace{-1mm}
	\begin{center}
    \resizebox{1.0\textwidth}{!}{%
		\begin{tabular}{c|c|c|c|c|c|c}
			\toprule[1.5pt]
			Method &data type
            & Dice $\uparrow$ & Jaccard $\uparrow$ & NSD $\uparrow$
            & ADB $\downarrow$  & HD95  $\downarrow$  \\
			\midrule[1.1pt]			
			2D U-Net~ (\cite{ronneberger2015u}) &2D
            & 63.98$\pm$19.72 & 49.63$\pm$18.20  & 72.61$\pm$23.01
            & 7.27$\pm$12.92  & 18.77$\pm$30.71
            \\ \hline
            DLA~(\cite{yu2018deep}) &2D
            & 64.82$\pm$18.66 & 50.23$\pm$16.74  & 73.29$\pm$21.21
            & 6.29$\pm$11.80  & 20.44$\pm$32.10
            \\ \hline
            U-Net++~(\cite{zhou2018unet++}) &2D
            & 66.01$\pm$19.12 & 51.70$\pm$17.35  & 72.23$\pm$21.77
            & 7.49$\pm$11.94& 23.93$\pm$33.81
            \\ \hline
            \hline
            3D U-Net~(\cite{cciccek20163d}) &3D
            & 65.91$\pm$22.55 & 52.60$\pm$20.82	 & 73.83$\pm$25.67
            & 10.42$\pm$22.53 & 27.00$\pm$46.60
            \\ \hline
            V-Net~(\cite{milletari2016v}) &3D
            & 65.89$\pm$20.10 & 51.89$\pm$18.71	& 70.69$\pm$24.82
            & 9.32$\pm$15.74  & 26.00$\pm$35.97
            \\ \hline
            nn-UNet~(\cite{isensee2019nnu}) &3D
            & 67.89$\pm$20.56	& 54.38$\pm$19.51 & 73.00$\pm$25.15	
            & 9.61$\pm$17.24	& 25.30$\pm$35.10
            \\ \hline \hline
            UA-MT~(\cite{yu2019uncertainty})  &3D
            & 69.32$\pm$19.07	& 55.60$\pm$17.85 & 74.73$\pm$23.20	
            & 8.22$\pm$16.52	& 24.60$\pm$36.69
            \\ \hline \hline
            \textbf{Our method~(DM${^2}$T-Net)}  &3D
            			
            & \textbf{72.59$\pm$18.55} & \textbf{59.42$\pm$17.09} & \textbf{80.44$\pm$20.19}
            & \textbf{4.45$\pm$8.06}   & \textbf{16.34$\pm$24.76}
            \\

			\bottomrule[1.5pt]
		\end{tabular}}
	\end{center}
	\vspace{-2.5mm}
\end{table*}

%% file: table-comparison-part2.tex
\begin{table*}[t]
	\caption{\wjc{The results (mean $\pm$ variance) of different segmentation methods on \textbf{MosMedData}. We use the bold fonts to highlight the best performance.}}
	\label{table:state-of-the-art-part2}
	\vspace{-1mm}
	\begin{center}
    \resizebox{1.0\textwidth}{!}{%
		\begin{tabular}{c|c|c|c|c|c|c}
			\toprule[1.5pt]
			Method &data type
            & Dice $\uparrow$ & Jaccard $\uparrow$ & NSD $\uparrow$
            & ADB $\downarrow$  & HD95  $\downarrow$  \\
			\midrule[1.1pt]			
			2D U-Net~ (\cite{ronneberger2015u}) &2D
            &53.47$\pm$22.39&39.47$\pm$19.75&77.94$\pm$23.52&6.45$\pm$15.51&17.61$\pm$22.40
            \\ \hline
            DLA~(\cite{yu2018deep}) &2D
            &55.71$\pm$18.90&40.91$\pm$17.75&80.34$\pm$18.20&5.08$\pm$7.75&22.01$\pm$28.64
            \\ \hline
            U-Net++~(\cite{zhou2018unet++}) &2D
            &56.03$\pm$22.46&42.05$\pm$20.39&77.71$\pm$22.71&5.51$\pm$6.94&22.05$\pm$28.17
            \\ \hline
            \hline
            3D U-Net~(\cite{cciccek20163d}) &3D
            &54.42$\pm$23.51&40.69$\pm$20.70&75.31$\pm$27.09&8.21$\pm$14.43&21.35$\pm$27.39
            \\ \hline
            V-Net~(\cite{milletari2016v}) &3D
            &48.95$\pm$19.76&34.61$\pm$17.03&71.88$\pm$20.39&5.16$\pm$8.19&21.20$\pm$18.23
            \\ \hline
            nn-UNet~(\cite{isensee2019nnu}) &3D
            &56.30$\pm$23.55&42.62$\pm$21.31&76.45$\pm$27.16&9.26$\pm$18.25&20.59$\pm$29.97
            \\ \hline \hline
            UA-MT~(\cite{yu2019uncertainty})  &3D
            &57.31$\pm$20.53&42.87$\pm$19.05&78.55$\pm$21.42&6.89$\pm$13.90&20.68$\pm$26.53
            \\ \hline \hline
            \textbf{Our method~(DM${^2}$T-Net)}  &3D
            &\textbf{60.19$\pm$19.22}&\textbf{45.56$\pm$18.44}
            &\textbf{80.95$\pm$20.99}&\textbf{6.55$\pm$13.79}
            &\textbf{21.11$\pm$27.74}\\
			\bottomrule[1.5pt]
		\end{tabular}}
	\end{center}
	\vspace{-2.5mm}
\end{table*}

%% file: experiments.tex
\section{Results and Discussions}
\label{sec:experiments}


\input{result-figure-part1}

\subsection{Evaluation Dataset and Metric}

\noindent
\textbf{Evaluation dataset.}
For evaluation, we build  a  new  CT  volume  segmentation dataset with 11 unlabeled data and 20 labeled data.
The labeled data is collected from COVID-19 3D CT dataset~\cite{ma2020towards}, which provides 20 COVID-19 CT volume data with pixel-level lung infection masks\wjc{, named COVID-19-P20}.
\wjc{The infections are firstly delineated by junior annotators with 1-5 years experience, then refined by two radiologists with 5-10 years experience, and finally all the annotations were verified and refined by a senior radiologist with more than 10 years experience in chest radiology.}
According to~\cite{ma2020towards}, the last 10 scans have been adjusted to the lung window [-1250, 250], and then normalized to [0, 255].
Meanwhile, we adjust the first 10 scans to the lung window [-1000,400], and the intensity values are normalized to [0,1].
Then we also collect 11 3D lung CT scans from 11 confirmed COVID-19 cases (8 female and 3 male), as the unlabeled data.
These unlabeled data is captured from Philips in Zhongshan Hospital, affiliated to Xiamen University.
Similar to the labeled dataset~\cite{ma2020towards}, we also adjust these 11 scans to the lung window [-1000, 400], and then normalized to [0, 1] for training.

Here, we conduct a two-fold evaluation on 20 labeled data. Specifically, COVID-19 3D CT dataset~\cite{ma2020towards} contains 10 CT volumes from Coronacases and 10 volumes from Radiopaedia. We randomly select $5$ volumes from Coronacases and randomly select $5$ volumes from Radiopaedia to form the first fold, and the remaining 10 scans are for the second fold. Then one fold combined with the unlabeled data is used as training set, while the rest fold is utilized as the test set.
Finally, we compute the mean$\pm$variance of two-fold segmentation results of each method for comparisons.

\wjc{We extensively build the experiment on a large public dataset, MosMedData~\cite{morozov2020mosmeddata}, which has 50 labeled CT volumes and 806 unlabeled CT volumes. 
The infections are annotated by the experts of Research and Practical Clinical Center for Diagnostics and Telemedicine Technologies of the Moscow Health Care Department. During the annotation for every given image ground-glass opacifications and regions of consolidation are selected as positive (white) pixels on the corresponding binary pixel mask.
Same pre-processing as COVID-19-P20 is adopted here and five-fold evaluation is conducted on the labeled data.}

\noindent
\textbf{Evaluation metric.}
We employ five widely-used metrics to quantitatively evaluate the COVID-19 lung infection segmentation performances, including $Dice$ coefficient, $Jaccard$ coefficient, normalized surface dice (NSD), average distance of boundaries (ADB), and  Hausdorff distance of boundaries (95$^{th}$ percentile; HD95).
In general, a better segmentation performance shall have higher $Dice$, $Jaccard$, and NSD scores, as well as lower ADB and HD95 scores.
\wjc{It is noteworthy that all the metrics in this paper are calculated in volume-wise, which is more meaningful for clinics and convincing for the assessment of 3D segmentation.}

$Dice$ and  $Jaccard$ coefficients compute the region-based similarity of the predicted segmentation result $P$ and the ground truth $G$:
\begin{equation}
Dice = \dfrac{2 \cdot \left\vert P \cap G \right\vert}{\left\vert P \right\vert + \left\vert G \right\vert}, \; Jaccard = \dfrac{ \left\vert P \cap G \right\vert}{\left\vert P \cup G \right\vert},
\end{equation}
where $\left\vert P \cap G \right\vert$ denotes the number of voxels in the intersection area of $P$ and $G$;
$\left\vert P \cup G \right\vert$ is the number of voxels in the union area of $P$ and $G$;
$\left\vert P \right\vert$ and $\left\vert G \right\vert$ are the number of voxels in the region $P$ and the region $G$, respectively.

NSD evaluates how close the segmentation and ground truth surfaces are to each
other at a specified tolerance, defined as:
\begin{equation}
NSD = \frac{\left\vert \partial G \cap B_{\partial  P}^{(\tau)} \right\vert + \left\vert \partial P \cap B_{\partial  G}^{(\tau)} \right\vert }{\left\vert \partial G \right\vert + \left\vert \partial P \right\vert} \ ,
\end{equation}
where $B_{\partial  P}^{(\tau)}$ and $B_{\partial  G}^{(\tau)}$ denote the border region of segmentation surface and ground truth, and they are: $B_{\partial  P}^{(\tau)}=$$\{u \in R^3 | \exists \widetilde{u} \in \partial P, \| u - \widetilde{u} \| \leq  \tau\}$, and $B_{\partial  G}^{(\tau)}=$$\{v \in R^3 | \exists \widetilde{v} \in \partial G, \| v - \widetilde{v} \| \leq  \tau\}$. And tolerance $\tau$ is empirically set as 1 mm and 3 mm for lung segmentation and infection segmentation, respectively.

ADB and HD estimate the surface distance between the predicted segmentation result and the manual ground truth:
\begin{equation}
    \begin{aligned}
        & ADB = \frac{1}{2} \Bigg\{ \frac{\sum_{v_i\in \Phi_{P}} h(v_i, \Phi_{G})}{\left\vert G \right\vert } + \frac{\sum_{v_j\in \Phi_{G}} h(v_j, \Phi_{P})}{\left\vert P \right\vert } \Bigg\} \ , \\
        & HD =  \max\left( \max_{v_i\in \Phi_{P}} h\left(v_i, \Phi_{G} \right), \max_{v_j\in \Phi_{G}} h(v_j, \Phi_{P}) \right) \ , \\
        & h(v_i, \Phi_{G}) = \min_{v_j \in \Phi_{G}} dist(v_i, v_j) \ , \\
         & h(v_j, \Phi_{P}) = \min_{v_i \in \Phi_{P}} dist(v_j, v_i) \ , \\
    \end{aligned}
\end{equation}
where $\Phi_{P}$ and $\Phi_{G}$ denote the surface of the prediction segmentation and ground truth, respectively. $v_i$ is a vertex of $\Phi_{P}$ and $v_j$ is a vertex of $\Phi_{G}$.
$dist(v_i, v_j)$ is the Euclidean distance between the vertex $v_i$ and the vertex $v_j$.
Apparently, ADB counts the average surface distance of the predicted segmentation and the ground truth surfaces.
HD computes the maximum distance between two segmentation surfaces, and HD95 is a modified HD by using the 95\% percentile instead of the maximum distance (100\% percentile) in HD in order to eliminate the impact of a small subset of the outliers; see~\cite{yu2019uncertainty} for more details.

\input{result-figure-part2}

\subsection{Implementation}

\noindent
\textbf{Training parameters.}
All parameters of our network are initialized from scratch, without requiring any pre-trained weight.
We augment the training set using a random flipping in all directions and adding Gaussian noise with the noise intensity $\sigma$$=$$10$.
Adam is employed to optimize the whole network with an initial learning rate of $0.0003$ and 5,000 iterations.
\wjc{We randomly sample 3D blocks with a size of $160\times160\times64$ for COVID-19-P20 or $160\times160\times32$ for MosMedData from each training CT volume for training our network on a single TITAN RTX.}

The mini-batch size is $4$, which consists of $2$ labeled images and $2$ unlabeled images.
The model size of our network is 18.79 Mb (megabyte), and the training time is 23 hours.

\noindent
\textbf{Inference.}
In the testing stage, we adopt the sliding window with a $50\%$ overlapping rate to
continually crop a set of volumes with a size of \wjc{$160\times160\times64$ for COVID-19-P20 or $160\times160\times32$ for MosMedData}. Moreover, we feed these cropped volumes into the student network of developed DM$^2$T-Net to generate multiple segmentation masks. Finally, we obtain the final segmentation of our network by stitching these small segmentation
masks according to their crop positions.
The average inference time (including pre-processing time) is $2.12$ seconds for one volume.


\input{table-ablation-study}


\input{result-ablation-study}

\subsection{Comparison with the State-of-the-art Methods}

We compare our method against seven state-of-the-art segmentation methods, including
2D U-Net~\cite{ronneberger2015u}, U-Net++~\cite{zhou2018unet++},
DLA~\cite{yu2018deep},
3D U-Net~\cite{cciccek20163d}, V-Net~\cite{milletari2016v},
nn-UNet~\cite{isensee2019nnu}, and UA-MT~\cite{yu2019uncertainty}.
Among them, the first three segmentation methods are based on 2D images while the other four methods directly perform the segmentation on 3D volumes. And the last one (UA-MT) is a state-of-the-art 3D semi-supervised segmentation method, which presented an uncertainty-aware self-ensembling model.
To make the comparisons fair, we adopt the released code of compared methods and fine-tune the parameters to obtain their best segmentation results.

Table~\ref{table:state-of-the-art} reports the quantitative results of different methods \wjc{on COVID-19-P20}.
Apparently, the 3D deep-learning-based methods~\cite{cciccek20163d,milletari2016v,isensee2019nnu} have superior performance of five metrics scores than 2D-image-based CNNs (i.e., 2D U-Net~\cite{ronneberger2015u} and 2D U-Net++~\cite{zhou2018unet++}), since these 3D segmentation methods can learn more inter-slice relations among 3D volume and reduce the false predictions on 2D slice without any COVID lung infection, which usually happen in the segmentation results of 2D U-Net and 2D U-Net++.
Moreover, due to the additional unlabeled data in the training set, the semi-supervise method, UA-MT~\cite{yu2019uncertainty}, further outperforms all these 3D supervised techniques in terms of all the five metrics.
Compared to the best-performing existing
method (UA-MT), our method has 4.72\% improvement on Dice, 6.87\% improvement on Jaccard, 7.64\% improvement on Jaccard, and 45.86\% reduction on ADB, 33.58\% reduction on HD95, respectively.
It indicates that our network can more accurately detect COVID-19 infected lung regions than state-of-the-art methods from 3D CT scans.
\wjc{We extensively evaluate the effectiveness of our method on MosMedData. The results in Table~\ref{table:state-of-the-art-part2} show that our network has larger Dice, Jaccard, and NSD scores, as well as smaller ADB and HD95 scores than state-of-the-art methods. It further indicates that our network can more accurately segment COVID-19 infected regions from CT scans.
Moreover, compared to COVID-19-P20, we can find that our network and state-of-the-art methods suffer from a degraded performance on MosMedData  for all five metrics. The main reason is that the data in MosMedData is more challenging and the infected regions in MosMedData are smaller than COVID-19-P20, thereby increasing the segmentation difficulties.
On the other hand, we argue that there are two main reasons why 2D U-Net performs poorly on two datasets in our network. First, existing works computed the DSC value of 2D U-Net on 2D slices, while our work computes all five metrics (including DSC metric) on the whole 3D volume. Second, existing works have not tested the 2D U-Net model on slices without COVID-19 infections, and thus a large amount of false positives will not be involved for computing the DSC score. }

Figs.~\ref{fig:comparison_real_photos_part1} and~\ref{fig:comparison_real_photos_part2} visually compare the COVID-19 lung infection segmentation results produced by our network and compared methods.
Apparently, compared methods tend to include many non-infection regions or neglect parts of infection regions in their segmentation results, while our network predicts more accurate infection segmentation results.
For these challenging inputs with multiple infection regions and different infection region sizes in Fig.~\ref{fig:comparison_real_photos_part2}, our network can still better segment these infected regions than all the compared methods.
It further verifies the effectiveness of the developed dual multi-scale mean teacher framework in our work.
\wjc{From the perspective of clinical importance, our method brings obvious improvement on those small and challenging infections, which are even hard for junior radiologists to determine. Although these infections are too small to make a great difference on the statistics, successful segmentation of them has much larger significance in practice.
}

\subsection{Ablation Analysis}

Here, we provide several experiments to validate
the effectiveness of main components of our network, including the multiple dimensional scale mechanism, multi-scale supervised loss, and multi-scale consistency loss.

\noindent
\textbf{Effectiveness of multiple dimensional-scale down-sampling.}
First of all, we construct a basic model (denoted as ``basic'') by removing the teacher network from our method, the dimensional-scale down-sampling operations of the input volume, and the segmentation predictions $P_2$, $P_3$, $P_4$, and $P_5$; see Fig.~\ref{fig:arc}.
In this way, the basic model almost becomes the classical 3D U-Net model.
After that, we add the the multiple dimensional scale mechanism to ``basic'' by fusing features from multiple down-sampled volumes for building another model (denoted as ``basic+multi-dimensional-scale'') to evaluate the contribution of our multiple dimensional-scale mechanism.
As shown in Table~\ref{table:ablation-study}, ``basic+multi-dimensional-scale'' has a superior performance of five metrics over ``basic''.
It shows that fusing features from multiple down-sampled volumes enables our method to accurately identify COVID-19 lung infected regions.

\noindent
\textbf{Effectiveness of multi-scale supervised loss.}
We then investigate the importance of the multi-scale supervised loss.
To do so, we build a model (denoted as ``basic+dual-multiscale'') by removing the multi-scale consistency loss from our network.
Compared to ``basic+multi-dimensional-scale'', we predict additional four segmentation results from $P_2$, $P_3$, $P_4$, and $P_5$ in ``basic+dual-multiscale'', and thus formulate the multi-scale supervised loss (see Eq.~\eqref{Eq:total_supervised_loss}).
From the results shown in Table~\ref{table:ablation-study}, ``basic+dual-multiscale'' performs better than ``basic+multi-dimensional-scale''.
It demonstrates that aggregating the supervised losses from different CNN layers via a multi-scale supervised loss helps our method to better identify COVID-19 infected regions in our method.

\noindent
\textbf{Effectiveness of multi-scale consistency loss.}
We finally investigate the importance of the multi-scale consistency loss by constructing another two models with unlabeled data.
The first one (``semi-basic'') adds the unlabeled data and encourages the segmentation results of ``basic'' from the student network and the teacher network to be consistent.
The second model (``semi-multi-dimensional-scale'') is to produce the infection segmentation results via 'basic+multi-dimensional-scale' and regularize the segmentation results from the student and teacher network to be consistent.

Table~\ref{table:ablation-study} reports the quantitative results of ``semi-basic'', ``semi-multi-dimensional-scale'', and our method.
Apparently,  ``semi-basic'' can more accurately segment COVID-19 infected lung regions than ``basic'' due to its superior performance of all the five metrics.
It indicates that the additional consistency loss from the unlabeled data incurs a superior infection segmentation performance.
Then, as shown in Table~\ref{table:ablation-study}, ``semi-multi-dimensional-scale'' outperforms ``semi-basic'' in terms of all five metrics, demonstrating that aggregating CNN features from multiple dimensional-scaled inputs enables a more accurate supervised loss and a more accurate consistency loss, thereby improving the lung infection segmentation accuracy.
Moreover, compared to ``semi-multi-dimensional-scale'', our method has higher Dice, Jaccard, and NSD scores, as well as lower ADB and HD95 scores, which shows that computing the consistency loss from multi-scale predictions at CNN layers further boosts the segmentation performance of our method.

\noindent
\textbf{Visual comparisons.} Fig.~\ref{fig:ablation_study} visually compares the segmentation results produced by our method and the five constructed baseline networks of the ablation study experiment (see Table~\ref{table:ablation-study}).
By observing these segmentation results, we can easily find that our DM$^2$T-Net better segments the COVID-19 infections than all the five baseline networks.
It further proves the effectiveness of considering unlabeled data and dual multi-scale information within an end-to-end network in our work.

\begin{figure}
    \centering
    \begin{subfigure}{0.9\linewidth}
        \includegraphics[width=\textwidth]{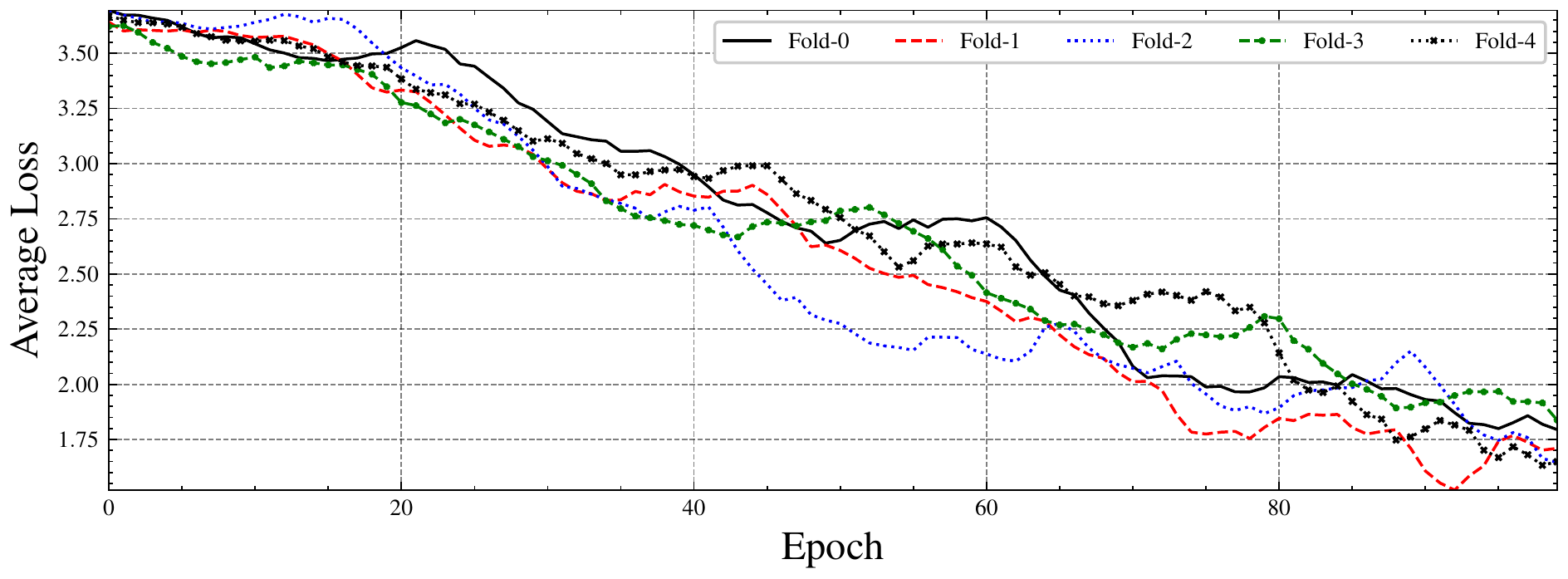}\vspace{-1mm}
        \caption{Training multi-scale supervised loss.}\vspace{1mm}
        \label{fig:sup_loss_curve}
    \end{subfigure}
    \begin{subfigure}{0.9\linewidth}
        \includegraphics[width=\textwidth]{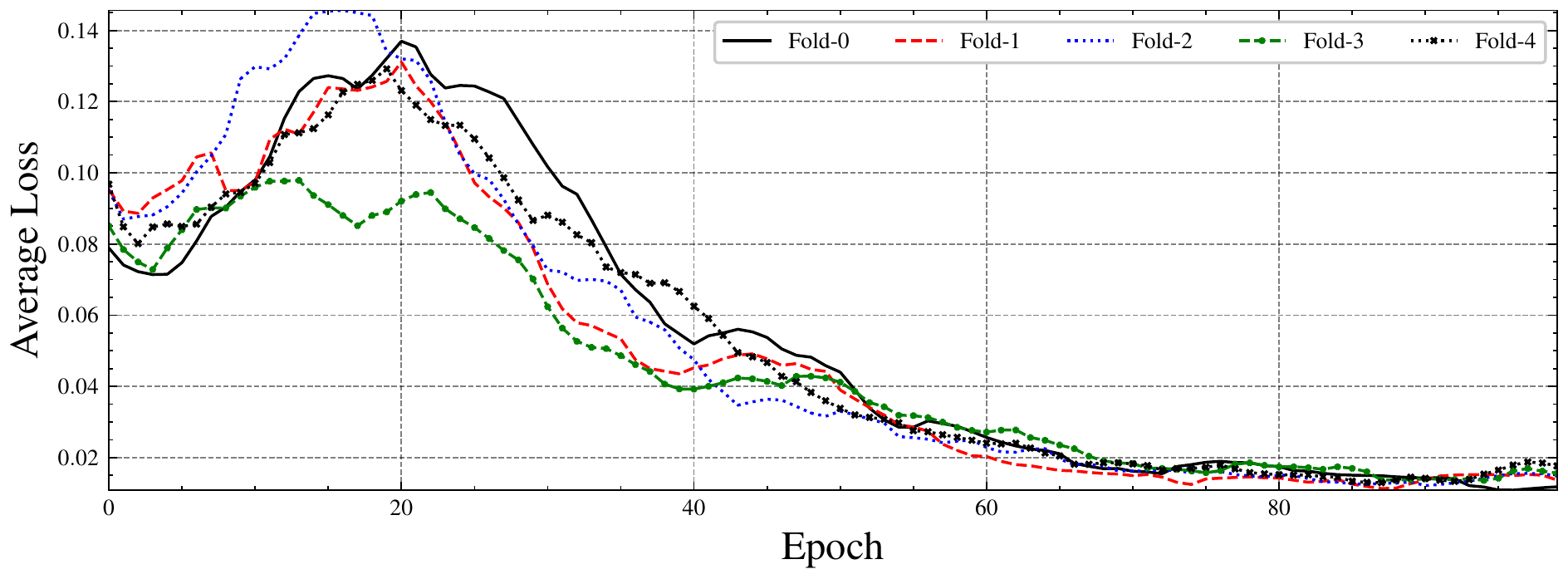}\vspace{-1mm}
        \caption{Training multi-scale consistency loss.}\vspace{1mm}
        \label{fig:consis_loss_curve}
    \end{subfigure}
    \vspace{1mm}
    \begin{subfigure}{0.9\linewidth}
        \includegraphics[width=\textwidth]{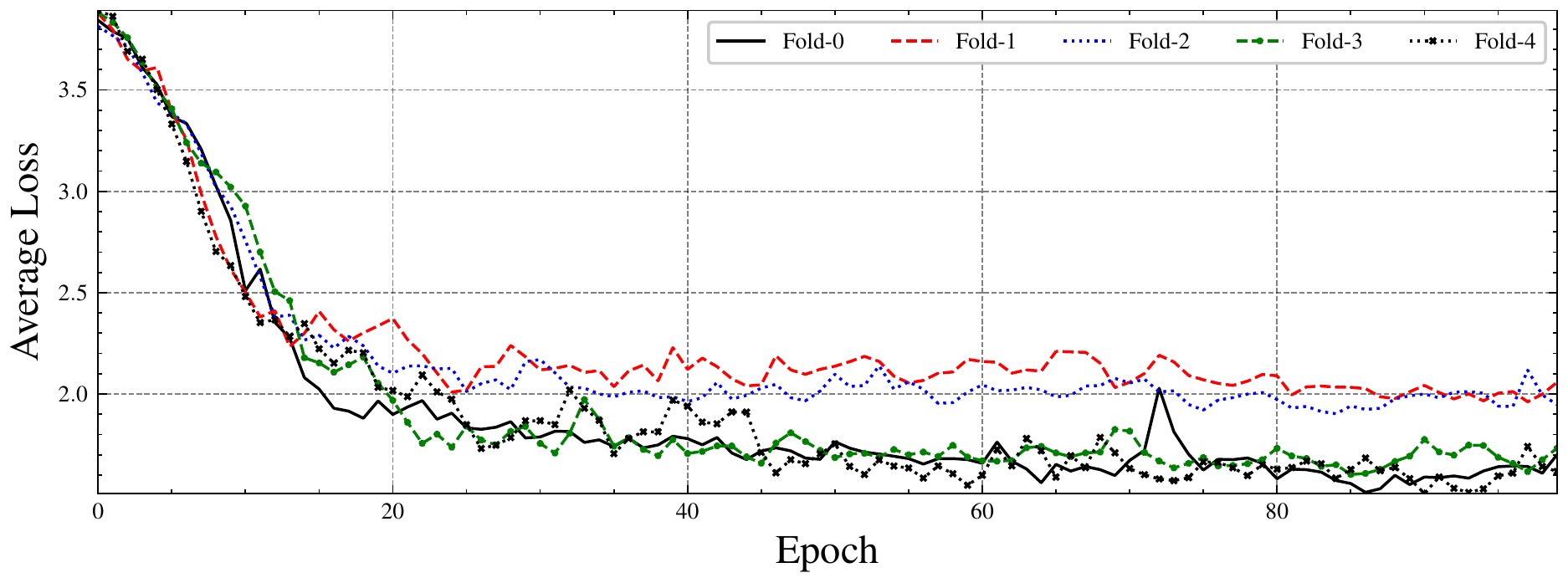}\vspace{-1mm}
        \caption{Testing multi-scale supervised loss.}
        \label{fig:test_loss_curve}
    \end{subfigure}
    \caption{Loss analysis during training our dual multi-scale mean teacher network \wjcfinal{on the COVID-19-P20 dataset}.}
    \label{fig:loss_curve}
\end{figure}

\subsection{Discussion}
\wjc{In this part, we provide the detailed analysis about (1) how the multi-scale consistency loss makes contribution to the DM${^2}$T-Net, including the loss curves and visual consistency between teacher and student networks, and (2) generalization analysis of the infection segmentation performance.}

\vspace{2mm}
\noindent
\wjc{\textbf{How does multi-scale consistency loss work?} Fig.~\ref{fig:loss_curve} presents the loss curves during semi-supervised learning process. It includes the multi-scale supervised segmentation loss (Fig.~\ref{fig:sup_loss_curve}) for labeled data and multi-scale consistency loss (Fig.~\ref{fig:consis_loss_curve}) for unlabeled data during training, as well as the testing multi-scale supervised loss (Fig.~\ref{fig:test_loss_curve}). 
It is observed that the consistency loss progressively increases at the first 20 epochs and then converges to a lower value after training 60 epochs. Since the teacher network's parameters are an average of consecutive student networks~\cite{tarvainen2017mean}, the difference at early epochs will be increasing and the averaging weight over large training steps tends to produce a more accurate supervision to the student network. Moreover, the supervised loss on the training set and the testing set decreases and then reaches stable values as the epoch number increases.}

\begin{figure}
    \centering
    \includegraphics[width=\linewidth]{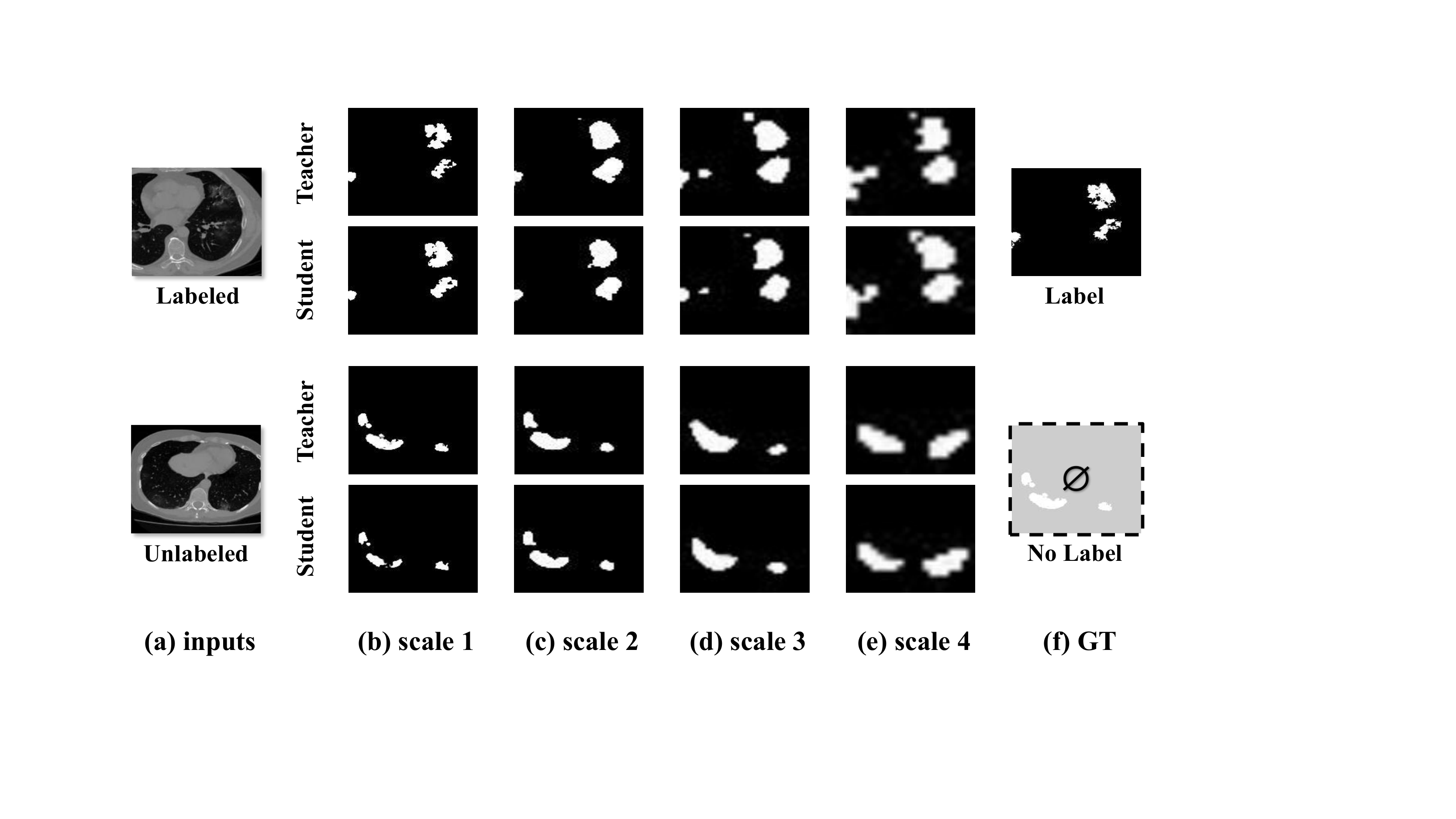}
    \caption{Segmentation results from the teacher and student networks at different scales \wjcfinal{on the COVID-19-P20 dataset}.}
    \label{fig:consis_result}
\end{figure}

\wjc{We additionally visualize all layers’ outputs (scales) from the teacher and student networks in Fig.~\ref{fig:consis_result}. First, considering the teacher network's parameters are the averaged result of the student network's, we can find that the teacher network's predictions are naturally more accurate than that of the student network, when compared to the ground truth of the labeled data. More importantly, combining the multi-scale labeled data loss and the multi-scale unlabeled data enables our network to produce good predictions for labeled data and unlabeled data. Hence, integrating the unlabeled data into the network training improves the COVID-19 infected region segmentation performance.}

\vspace{2mm}
\noindent
\wjc{\textbf{Generalization analysis}. We make the discussion about model generalization in Table~\ref{tab:gl}, where we use the model trained on one dataset to segment the infections of the other one dataset and assess the performance. From the quantitative results, we have the following observations: (1) due to the imaging variance and infection difference on two datasets, the segmentation performance of both 3D-based UA-MT and 2D-based U-Net++ decrease, but the 3D-based segmentation method have a better generalization capability than the 2D-based U-Net++ on both cross-dataset evaluation settings. The underlying reason is that the 3D-based method is able to capture more high-level information of lung infections among multiple image slices than the 2D-based segmentation performance. (2) More importantly, compared to the best existing 3D-based method (UA-MT), our method still outperforms it in terms of Dice and HD95, which demonstrates the generalized advancement of our network. We have added this experiment into Section IV.E of the revised manuscript.}

\input{table-generalization}

%% file: result-figure-part1.tex
\begin{figure*}[htbp]
	\centering
    \vspace*{0.5mm}
	\begin{subfigure}{0.092\textwidth}
		\includegraphics[width=\textwidth]{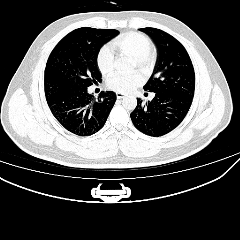}
	\end{subfigure}
	\begin{subfigure}{0.092\textwidth}
		\includegraphics[width=\textwidth]{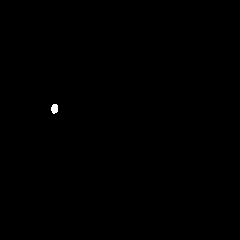}
	\end{subfigure}
	\begin{subfigure}{0.092\textwidth}
		\includegraphics[width=\textwidth]{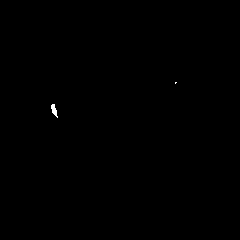}
	\end{subfigure}
	\begin{subfigure}{0.092\textwidth}
		\includegraphics[width=\textwidth]{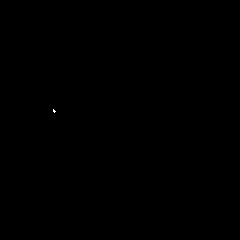}
	\end{subfigure}
	\begin{subfigure}{0.092\textwidth}
		\includegraphics[width=\textwidth]{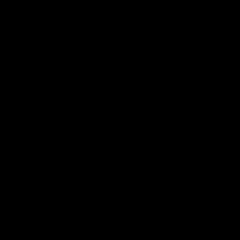}
	\end{subfigure}
	\begin{subfigure}{0.092\textwidth}
		\includegraphics[width=\textwidth]{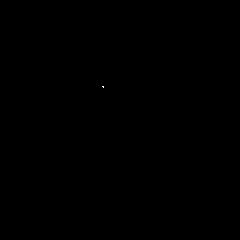}
	\end{subfigure}
	\begin{subfigure}{0.092\textwidth}
		\includegraphics[width=\textwidth]{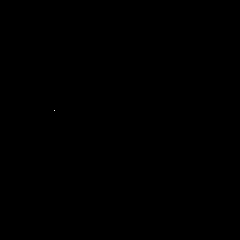}
	\end{subfigure}
	\begin{subfigure}{0.092\textwidth}
		\includegraphics[width=\textwidth]{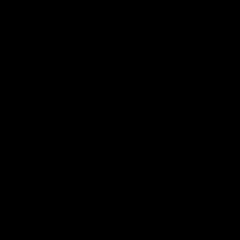}
	\end{subfigure}
    \begin{subfigure}{0.092\textwidth}
		\includegraphics[width=\textwidth]{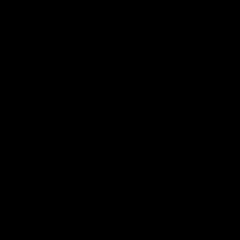}
	\end{subfigure}
	\begin{subfigure}{0.092\textwidth}
		\includegraphics[width=\textwidth]{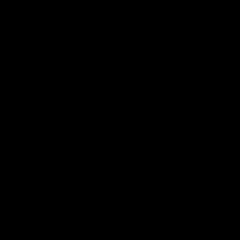}
	\end{subfigure}
	\ \\	
    \vspace*{0.5mm}
	\begin{subfigure}{0.092\textwidth}
		\includegraphics[width=\textwidth]{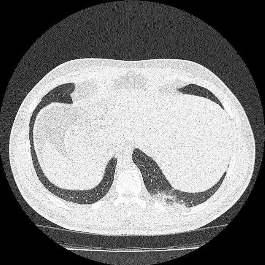}
	\end{subfigure}
	\begin{subfigure}{0.092\textwidth}
		\includegraphics[width=\textwidth]{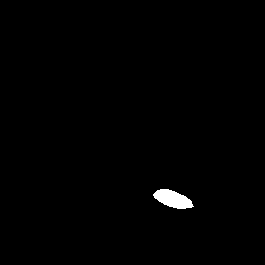}
	\end{subfigure}
	\begin{subfigure}{0.092\textwidth}
		\includegraphics[width=\textwidth]{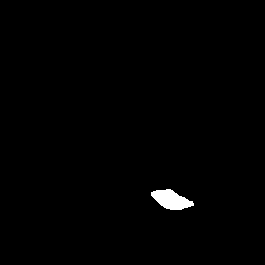}
	\end{subfigure}
	\begin{subfigure}{0.092\textwidth}
		\includegraphics[width=\textwidth]{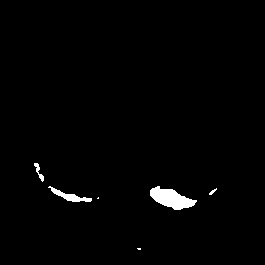}
	\end{subfigure}
	\begin{subfigure}{0.092\textwidth}
		\includegraphics[width=\textwidth]{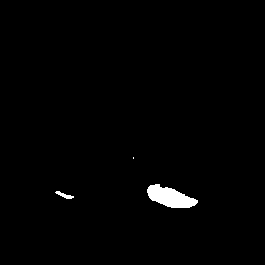}
	\end{subfigure}
	\begin{subfigure}{0.092\textwidth}
		\includegraphics[width=\textwidth]{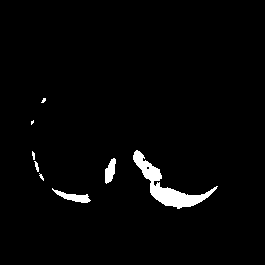}
	\end{subfigure}
	\begin{subfigure}{0.092\textwidth}
		\includegraphics[width=\textwidth]{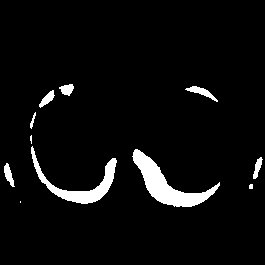}
	\end{subfigure}
	\begin{subfigure}{0.092\textwidth}
		\includegraphics[width=\textwidth]{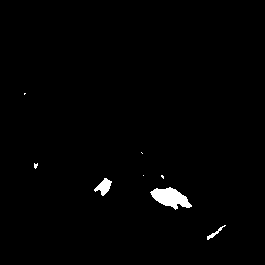}
	\end{subfigure}
    \begin{subfigure}{0.092\textwidth}
		\includegraphics[width=\textwidth]{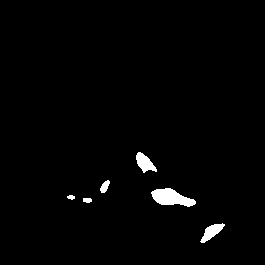}
	\end{subfigure}
	\begin{subfigure}{0.092\textwidth}
		\includegraphics[width=\textwidth]{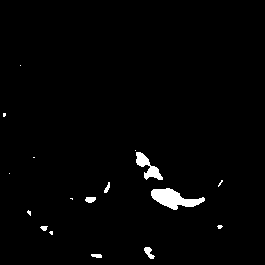}
	\end{subfigure}
	\ \\	
    \vspace*{0.5mm}
	\begin{subfigure}{0.092\textwidth}
		\includegraphics[width=\textwidth]{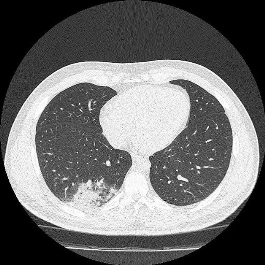}
	\end{subfigure}
	\begin{subfigure}{0.092\textwidth}
		\includegraphics[width=\textwidth]{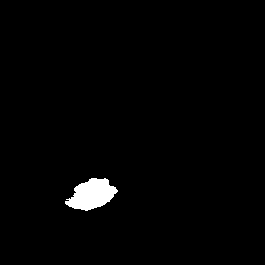}
	\end{subfigure}
	\begin{subfigure}{0.092\textwidth}
		\includegraphics[width=\textwidth]{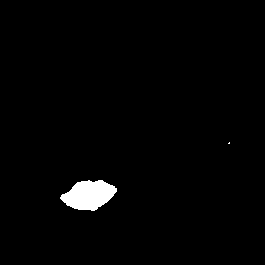}
	\end{subfigure}
	\begin{subfigure}{0.092\textwidth}
		\includegraphics[width=\textwidth]{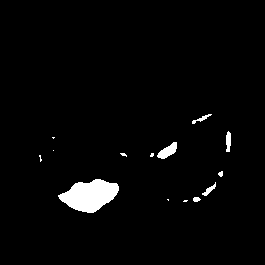}
	\end{subfigure}
	\begin{subfigure}{0.092\textwidth}
		\includegraphics[width=\textwidth]{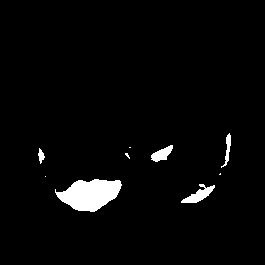}
	\end{subfigure}
	\begin{subfigure}{0.092\textwidth}
		\includegraphics[width=\textwidth]{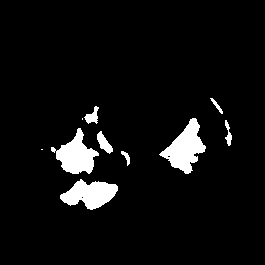}
	\end{subfigure}
	\begin{subfigure}{0.092\textwidth}
		\includegraphics[width=\textwidth]{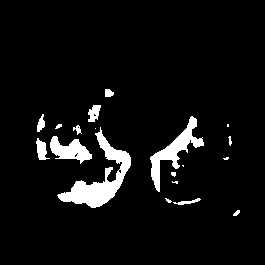}
	\end{subfigure}
	\begin{subfigure}{0.092\textwidth}
		\includegraphics[width=\textwidth]{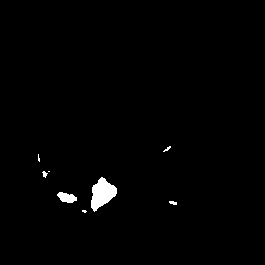}
	\end{subfigure}
    \begin{subfigure}{0.092\textwidth}
		\includegraphics[width=\textwidth]{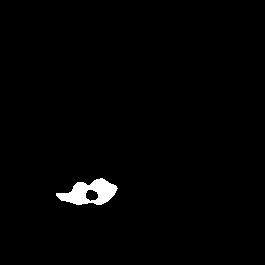}
	\end{subfigure}
	\begin{subfigure}{0.092\textwidth}
		\includegraphics[width=\textwidth]{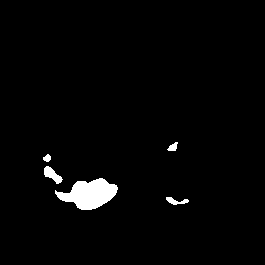}
	\end{subfigure}
	\ \\	
    \vspace*{0.5mm}
    \begin{subfigure}{0.092\textwidth}
		\includegraphics[width=\textwidth]{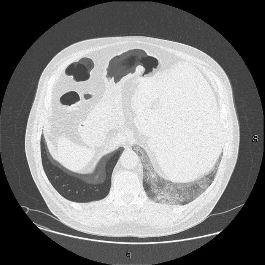}
	\end{subfigure}
	\begin{subfigure}{0.092\textwidth}
		\includegraphics[width=\textwidth]{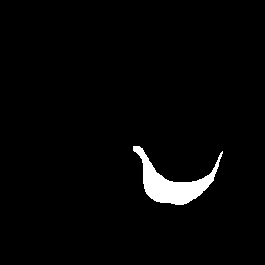}
	\end{subfigure}
	\begin{subfigure}{0.092\textwidth}
		\includegraphics[width=\textwidth]{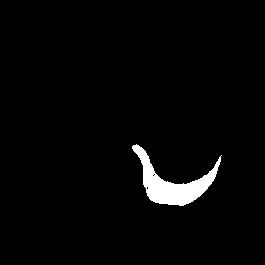}
	\end{subfigure}
	\begin{subfigure}{0.092\textwidth}
		\includegraphics[width=\textwidth]{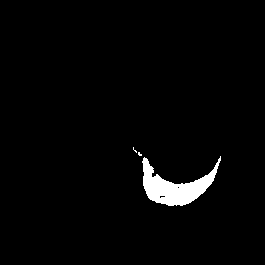}
	\end{subfigure}
	\begin{subfigure}{0.092\textwidth}
		\includegraphics[width=\textwidth]{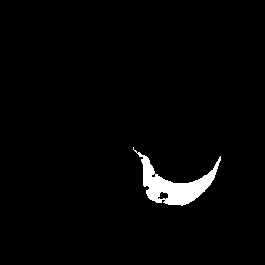}
	\end{subfigure}
	\begin{subfigure}{0.092\textwidth}
		\includegraphics[width=\textwidth]{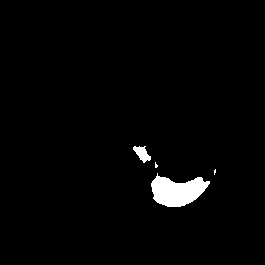}
	\end{subfigure}
	\begin{subfigure}{0.092\textwidth}
		\includegraphics[width=\textwidth]{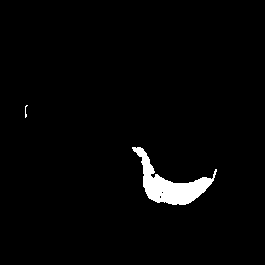}
	\end{subfigure}
	\begin{subfigure}{0.092\textwidth}
		\includegraphics[width=\textwidth]{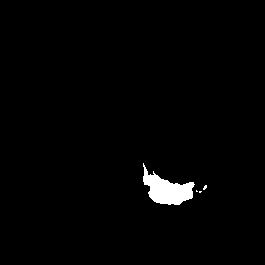}
	\end{subfigure}
    \begin{subfigure}{0.092\textwidth}
		\includegraphics[width=\textwidth]{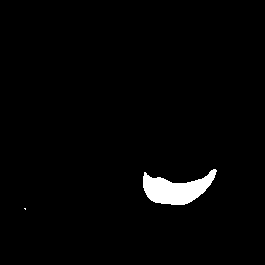}
	\end{subfigure}
    \begin{subfigure}{0.092\textwidth}
		\includegraphics[width=\textwidth]{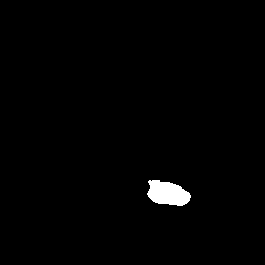}
	\end{subfigure}
	\ \\	
    \vspace*{0.5mm}
   \begin{subfigure}{0.092\textwidth}
   	\includegraphics[width=\textwidth]{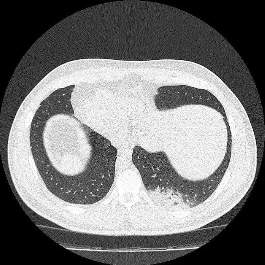}
    \vspace{-5.5mm} \caption{\footnotesize{inputs}}
   \end{subfigure}
   \begin{subfigure}{0.092\textwidth}
   	\includegraphics[width=\textwidth]{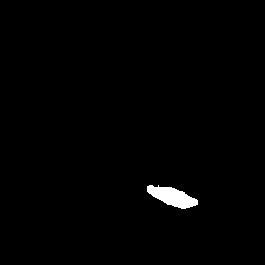}
   	\vspace{-5.5mm} \caption{\footnotesize{GT}}
   \end{subfigure}
   \begin{subfigure}{0.092\textwidth}
   	\includegraphics[width=\textwidth]{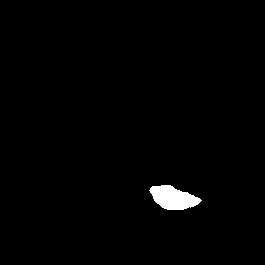}
    \vspace{-5.5mm} \caption{\footnotesize{ours}}
   \end{subfigure}
   \begin{subfigure}{0.092\textwidth}
   	\includegraphics[width=\textwidth]{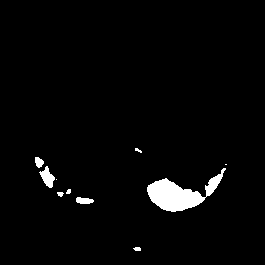}
    \vspace{-5.5mm} \caption{\scriptsize{UA-MT}}
   \end{subfigure}
   \begin{subfigure}{0.092\textwidth}
   	\includegraphics[width=\textwidth]{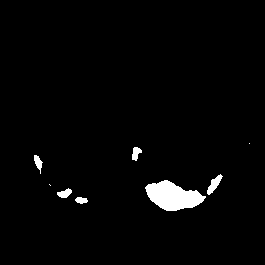}
   	\vspace{-5.5mm} \caption{\footnotesize{nn-UNet}}
   \end{subfigure}
   \begin{subfigure}{0.092\textwidth}
   	\includegraphics[width=\textwidth]{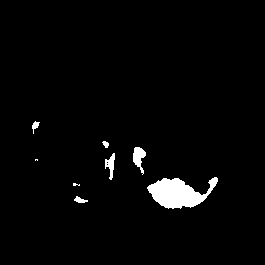}
    \vspace{-5.5mm} \caption{\footnotesize{V-Net}}
   \end{subfigure}
   \begin{subfigure}{0.092\textwidth}
   	\includegraphics[width=\textwidth]{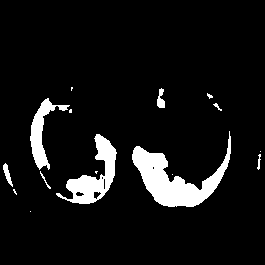}
    \vspace{-5.5mm} \caption{\footnotesize{3D U-Net}}
   \end{subfigure}
   \begin{subfigure}{0.092\textwidth}
   	\includegraphics[width=\textwidth]{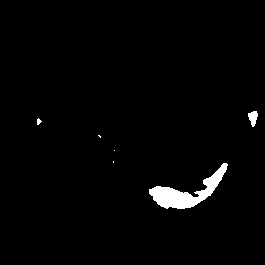}
    \vspace{-5.5mm} \caption{\scriptsize{DLA}}
   \end{subfigure}
   \begin{subfigure}{0.092\textwidth}
   	\includegraphics[width=\textwidth]{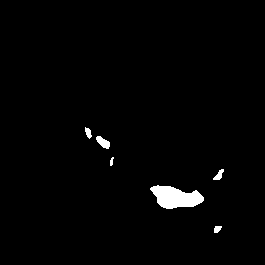}
    \vspace{-5.5mm} \caption{\footnotesize{U-Net++}}
   \end{subfigure}
   \begin{subfigure}{0.092\textwidth}
   	\includegraphics[width=\textwidth]{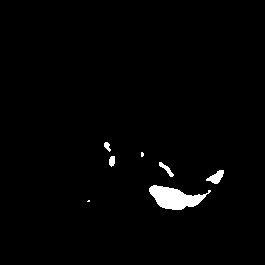}
    \vspace{-5.5mm} \caption{\scriptsize{2D U-Net}}
   \end{subfigure}
	\caption{Visual comparison of segmentation results produced by different methods \wjcfinal{on the COVID-19-P20 dataset}.
    (a) Input images;
    (b) Ground truths (denoted as 'GT');
    (c)-(h) segmentation results predicted by our method, UA-MT~\cite{yu2019uncertainty}, nn-UNet~\cite{isensee2019nnu},
    3D U-Net~\cite{cciccek20163d}, DLA~\cite{yu2018deep}, U-Net++~\cite{zhou2018unet++}, and 2D U-Net~\cite{ronneberger2015u}.
	Apparently, our network can more accurately identify COVID-19 lung infected regions than other methods.
    }
	\label{fig:comparison_real_photos_part1}
	
\end{figure*}

%% file: result-figure-part2.tex
\begin{figure*}[t]
	\centering
    \vspace*{0.5mm}
	\begin{subfigure}{0.092\textwidth}
		\includegraphics[width=\textwidth]{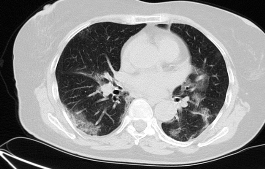}
	\end{subfigure}
	\begin{subfigure}{0.092\textwidth}
		\includegraphics[width=\textwidth]{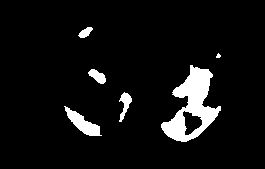}
	\end{subfigure}
	\begin{subfigure}{0.092\textwidth}
		\includegraphics[width=\textwidth]{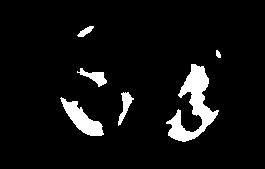}
	\end{subfigure}
	\begin{subfigure}{0.092\textwidth}
		\includegraphics[width=\textwidth]{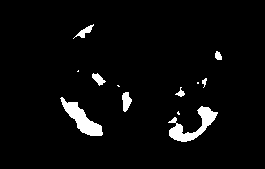}
	\end{subfigure}
	\begin{subfigure}{0.092\textwidth}
		\includegraphics[width=\textwidth]{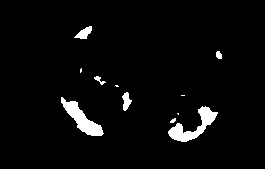}
	\end{subfigure}
	\begin{subfigure}{0.092\textwidth}
		\includegraphics[width=\textwidth]{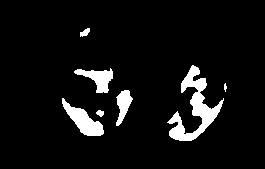}
	\end{subfigure}
	\begin{subfigure}{0.092\textwidth}
		\includegraphics[width=\textwidth]{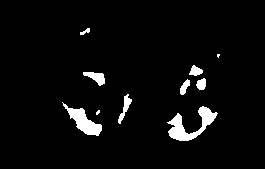}
	\end{subfigure}
	\begin{subfigure}{0.092\textwidth}
		\includegraphics[width=\textwidth]{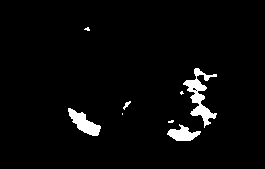}
	\end{subfigure}
    \begin{subfigure}{0.092\textwidth}
		\includegraphics[width=\textwidth]{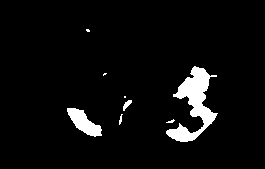}
	\end{subfigure}
	\begin{subfigure}{0.092\textwidth}
		\includegraphics[width=\textwidth]{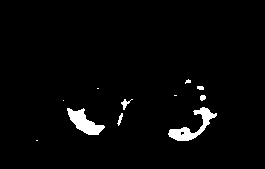}
	\end{subfigure}
	\ \\	
    \vspace*{0.5mm}
	\begin{subfigure}{0.092\textwidth}
		\includegraphics[width=\textwidth]{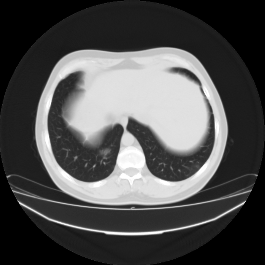}
	\end{subfigure}
	\begin{subfigure}{0.092\textwidth}
		\includegraphics[width=\textwidth]{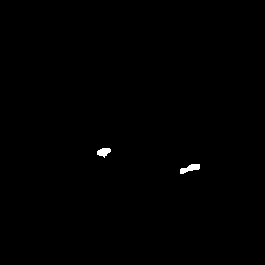}
	\end{subfigure}
	\begin{subfigure}{0.092\textwidth}
		\includegraphics[width=\textwidth]{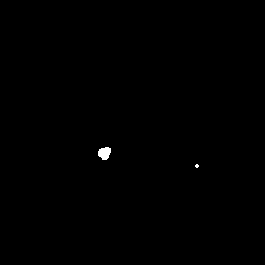}
	\end{subfigure}
	\begin{subfigure}{0.092\textwidth}
		\includegraphics[width=\textwidth]{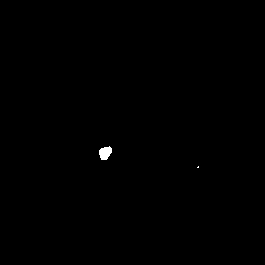}
	\end{subfigure}
	\begin{subfigure}{0.092\textwidth}
		\includegraphics[width=\textwidth]{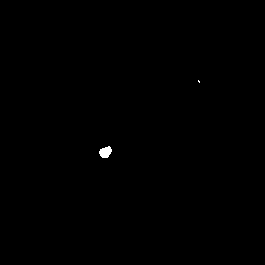}
	\end{subfigure}
	\begin{subfigure}{0.092\textwidth}
		\includegraphics[width=\textwidth]{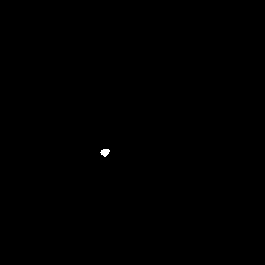}
	\end{subfigure}
	\begin{subfigure}{0.092\textwidth}
		\includegraphics[width=\textwidth]{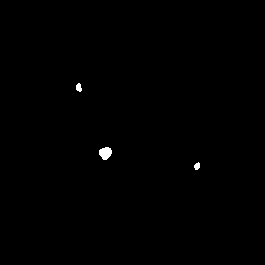}
	\end{subfigure}
	\begin{subfigure}{0.092\textwidth}
		\includegraphics[width=\textwidth]{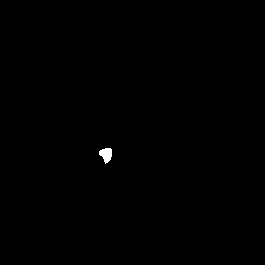}
	\end{subfigure}
    \begin{subfigure}{0.092\textwidth}
		\includegraphics[width=\textwidth]{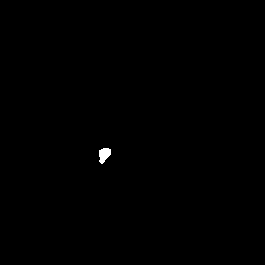}
	\end{subfigure}
	\begin{subfigure}{0.092\textwidth}
		\includegraphics[width=\textwidth]{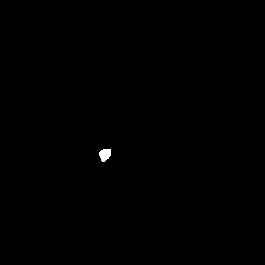}
	\end{subfigure}
	\ \\	
    \vspace*{0.5mm}
	\begin{subfigure}{0.092\textwidth}
		\includegraphics[width=\textwidth]{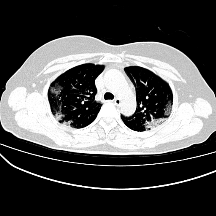}
	\end{subfigure}
	\begin{subfigure}{0.092\textwidth}
		\includegraphics[width=\textwidth]{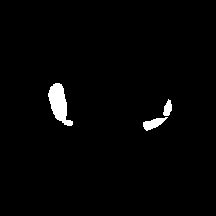}
	\end{subfigure}
	\begin{subfigure}{0.092\textwidth}
		\includegraphics[width=\textwidth]{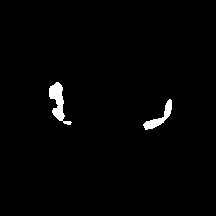}
	\end{subfigure}
	\begin{subfigure}{0.092\textwidth}
		\includegraphics[width=\textwidth]{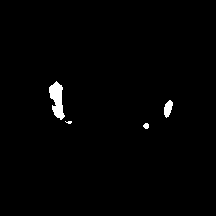}
	\end{subfigure}
	\begin{subfigure}{0.092\textwidth}
		\includegraphics[width=\textwidth]{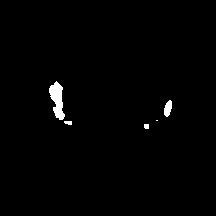}
	\end{subfigure}
	\begin{subfigure}{0.092\textwidth}
		\includegraphics[width=\textwidth]{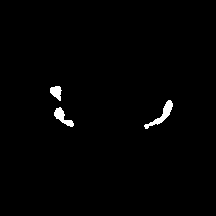}
	\end{subfigure}
	\begin{subfigure}{0.092\textwidth}
		\includegraphics[width=\textwidth]{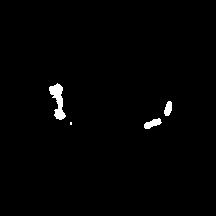}
	\end{subfigure}
	\begin{subfigure}{0.092\textwidth}
		\includegraphics[width=\textwidth]{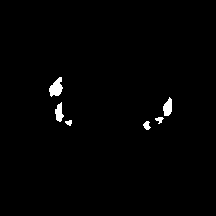}
	\end{subfigure}
    \begin{subfigure}{0.092\textwidth}
		\includegraphics[width=\textwidth]{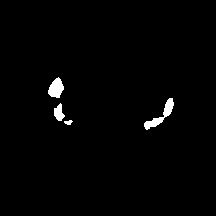}
	\end{subfigure}
	\begin{subfigure}{0.092\textwidth}
		\includegraphics[width=\textwidth]{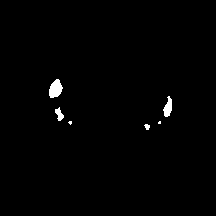}
	\end{subfigure}
	\ \\	
    \vspace*{0.5mm}
	\begin{subfigure}{0.092\textwidth}
		\includegraphics[width=\textwidth]{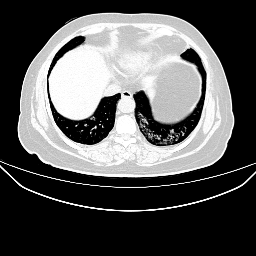}
	\end{subfigure}
	\begin{subfigure}{0.092\textwidth}
		\includegraphics[width=\textwidth]{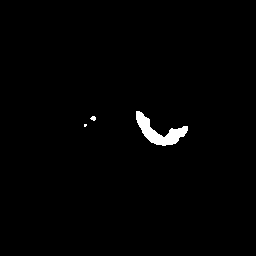}
	\end{subfigure}
	\begin{subfigure}{0.092\textwidth}
		\includegraphics[width=\textwidth]{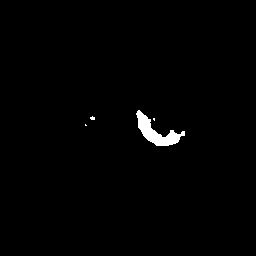}
	\end{subfigure}
	\begin{subfigure}{0.092\textwidth}
		\includegraphics[width=\textwidth]{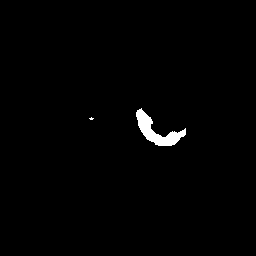}
	\end{subfigure}
	\begin{subfigure}{0.092\textwidth}
		\includegraphics[width=\textwidth]{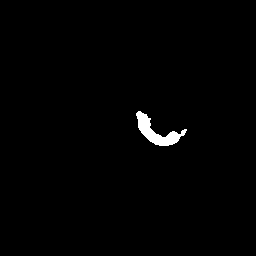}
	\end{subfigure}
	\begin{subfigure}{0.092\textwidth}
		\includegraphics[width=\textwidth]{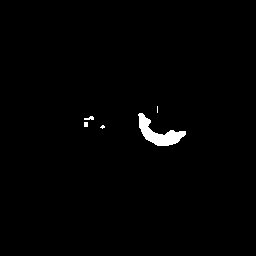}
	\end{subfigure}
	\begin{subfigure}{0.092\textwidth}
		\includegraphics[width=\textwidth]{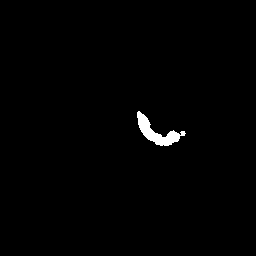}
	\end{subfigure}
	\begin{subfigure}{0.092\textwidth}
		\includegraphics[width=\textwidth]{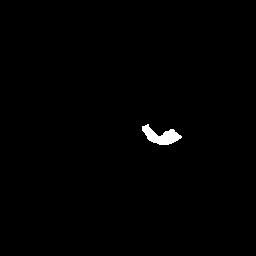}
	\end{subfigure}
    \begin{subfigure}{0.092\textwidth}
		\includegraphics[width=\textwidth]{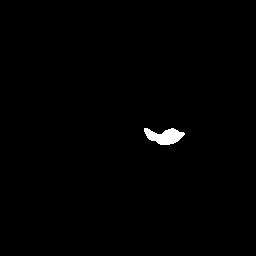}
	\end{subfigure}
	\begin{subfigure}{0.092\textwidth}
		\includegraphics[width=\textwidth]{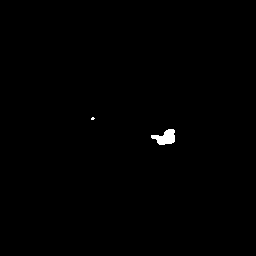}
	\end{subfigure}
	\ \\	
    \vspace*{0.5mm}
	\begin{subfigure}{0.092\textwidth}
		\includegraphics[width=\textwidth]{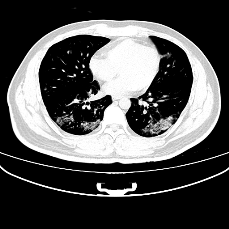}
	\end{subfigure}
	\begin{subfigure}{0.092\textwidth}
		\includegraphics[width=\textwidth]{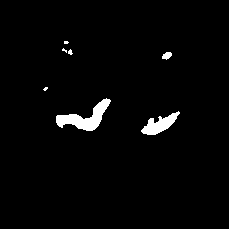}
	\end{subfigure}
	\begin{subfigure}{0.092\textwidth}
		\includegraphics[width=\textwidth]{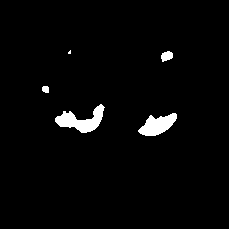}
	\end{subfigure}
	\begin{subfigure}{0.092\textwidth}
		\includegraphics[width=\textwidth]{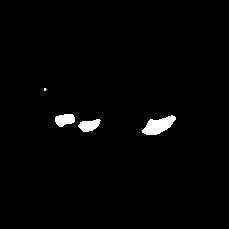}
	\end{subfigure}
	\begin{subfigure}{0.092\textwidth}
		\includegraphics[width=\textwidth]{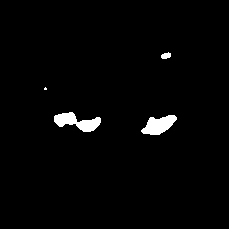}
	\end{subfigure}
	\begin{subfigure}{0.092\textwidth}
		\includegraphics[width=\textwidth]{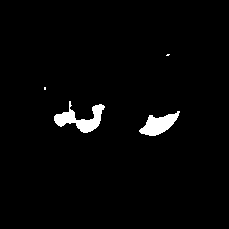}
	\end{subfigure}
	\begin{subfigure}{0.092\textwidth}
		\includegraphics[width=\textwidth]{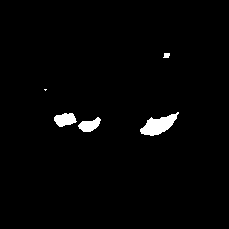}
	\end{subfigure}
	\begin{subfigure}{0.092\textwidth}
		\includegraphics[width=\textwidth]{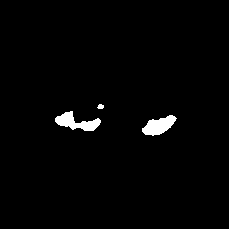}
	\end{subfigure}
    \begin{subfigure}{0.092\textwidth}
		\includegraphics[width=\textwidth]{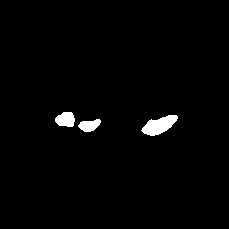}
	\end{subfigure}
	\begin{subfigure}{0.092\textwidth}
		\includegraphics[width=\textwidth]{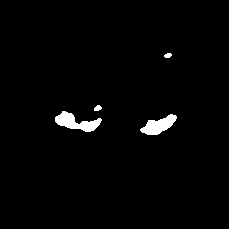}
	\end{subfigure}
	\ \\
    \vspace*{0.5mm}
	\begin{subfigure}{0.092\textwidth}
		\includegraphics[width=\textwidth]{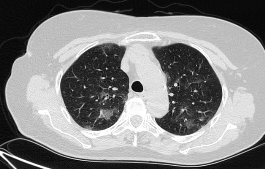}
	\end{subfigure}
	\begin{subfigure}{0.092\textwidth}
		\includegraphics[width=\textwidth]{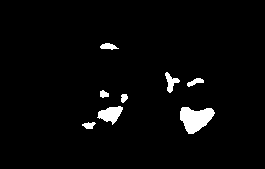}
	\end{subfigure}
	\begin{subfigure}{0.092\textwidth}
		\includegraphics[width=\textwidth]{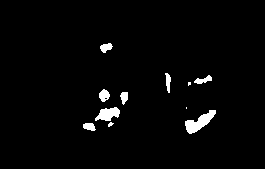}
	\end{subfigure}
	\begin{subfigure}{0.092\textwidth}
		\includegraphics[width=\textwidth]{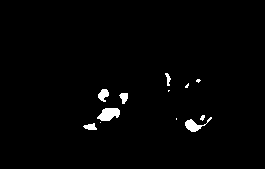}
	\end{subfigure}
	\begin{subfigure}{0.092\textwidth}
		\includegraphics[width=\textwidth]{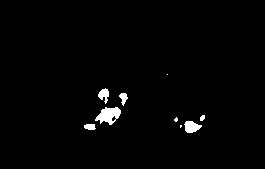}
	\end{subfigure}
	\begin{subfigure}{0.092\textwidth}
		\includegraphics[width=\textwidth]{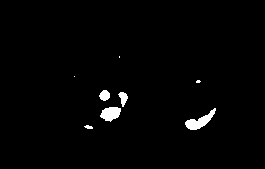}
	\end{subfigure}
	\begin{subfigure}{0.092\textwidth}
		\includegraphics[width=\textwidth]{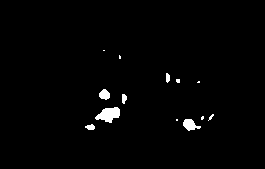}
	\end{subfigure}
	\begin{subfigure}{0.092\textwidth}
		\includegraphics[width=\textwidth]{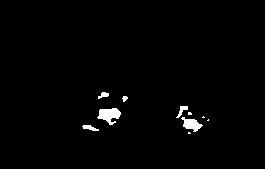}
	\end{subfigure}
    \begin{subfigure}{0.092\textwidth}
		\includegraphics[width=\textwidth]{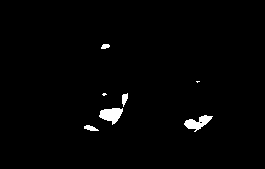}
	\end{subfigure}
	\begin{subfigure}{0.092\textwidth}
		\includegraphics[width=\textwidth]{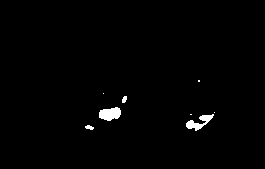}
	\end{subfigure}
	\ \\	
    \vspace*{0.5mm}
   \begin{subfigure}{0.092\textwidth}
   	\includegraphics[width=\textwidth]{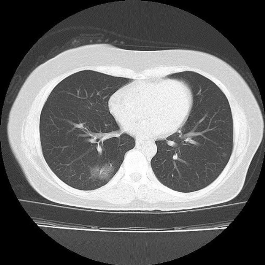}
    \vspace{-5.5mm} \caption{\footnotesize{inputs}}
   \end{subfigure}
   \begin{subfigure}{0.092\textwidth}
   	\includegraphics[width=\textwidth]{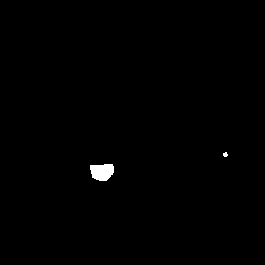}
   	\vspace{-5.5mm} \caption{\footnotesize{GT}}
   \end{subfigure}
   \begin{subfigure}{0.092\textwidth}
   	\includegraphics[width=\textwidth]{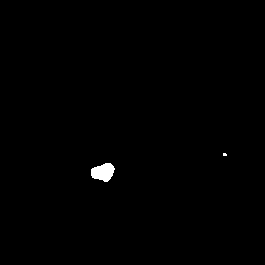}
    \vspace{-5.5mm} \caption{\footnotesize{ours}}
   \end{subfigure}
   \begin{subfigure}{0.092\textwidth}
   	\includegraphics[width=\textwidth]{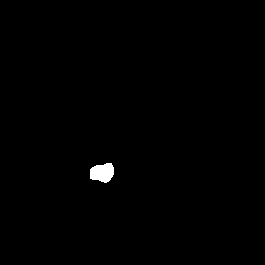}
    \vspace{-5.5mm} \caption{\scriptsize{UA-MT}}
   \end{subfigure}
   \begin{subfigure}{0.092\textwidth}
   	\includegraphics[width=\textwidth]{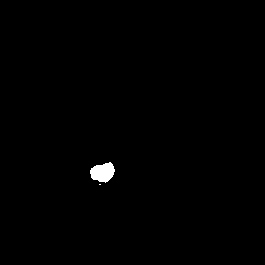}
   	\vspace{-5.5mm} \caption{\footnotesize{nn-UNet}}
   \end{subfigure}
   \begin{subfigure}{0.092\textwidth}
   	\includegraphics[width=\textwidth]{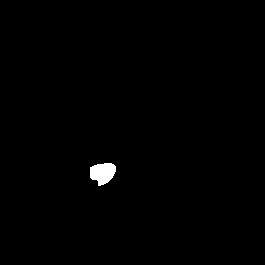}
    \vspace{-5.5mm} \caption{\footnotesize{V-Net}}
   \end{subfigure}
   \begin{subfigure}{0.092\textwidth}
   	\includegraphics[width=\textwidth]{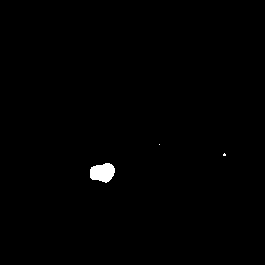}
    \vspace{-5.5mm} \caption{\footnotesize{3D U-Net}}
   \end{subfigure}
   \begin{subfigure}{0.092\textwidth}
   	\includegraphics[width=\textwidth]{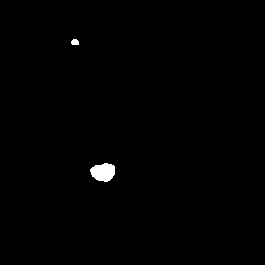}
    \vspace{-5.5mm} \caption{\scriptsize{DLA}}
   \end{subfigure}
   \begin{subfigure}{0.092\textwidth}
   	\includegraphics[width=\textwidth]{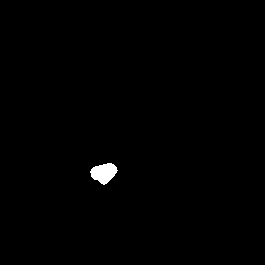}
    \vspace{-5.5mm} \caption{\footnotesize{U-Net++}}
   \end{subfigure}
   \begin{subfigure}{0.092\textwidth}
   	\includegraphics[width=\textwidth]{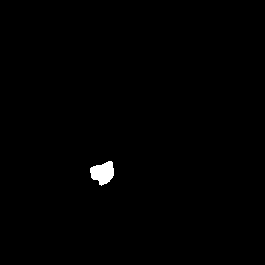}
    \vspace{-5.5mm} \caption{\scriptsize{2D U-Net}}
   \end{subfigure}
	\caption{Visual comparison of segmentation results produced by different methods \wjcfinal{on the COVID-19-P20 dataset} (continued from Fig.~\ref{fig:comparison_real_photos_part1}).
    (a) Input images with multiple infected regions;
    (b) Ground truths (denoted as 'GT');
    (c)-(h) segmentation results predicted by our method, UA-MT~\cite{yu2019uncertainty}, nn-UNet~\cite{isensee2019nnu},
    3D U-Net~\cite{cciccek20163d}, DLA~\cite{yu2018deep}, U-Net++~\cite{zhou2018unet++}, and 2D U-Net~\cite{ronneberger2015u}.
	Apparently, our network can more accurately identify COVID-19 lung infected regions than other methods.
    }
	\label{fig:comparison_real_photos_part2}
	
\end{figure*}

%% file: table-ablation-study.tex
\begin{table*}[t]
	\caption{The results (mean $\pm$ variance) of different ablation study experiments \wjcfinal{on the COVID-19-P20 dataset}. We use the bold fonts to highlight the best performance.}
	\label{table:ablation-study}
	\vspace{-1mm}
	\begin{center}
    \resizebox{1.0\textwidth}{!}{%
		\begin{tabular}{c|c|c|c|c|c|c|c|c}
			\toprule[1.5pt]
			Method & multiple-dimension-scale &dual multi-scale &semi-supervised & Dice $\uparrow$ & Jaccard $\uparrow$  & NSD 
            $\uparrow$  & ADB $\downarrow$  & HD95  $\downarrow$  \\
			\midrule[1.1pt]			
			basic & $\times$ & $\times$ & $\times$
            & 67.89$\pm$20.56	& 54.38$\pm$19.51 & 73.00$\pm$25.15	
            & 9.61$\pm$17.24	& 25.30$\pm$35.10
            \\ \hline
            basic+multi-dimensional-scale & $\surd$ & $\times$ & $\times$
            & 70.86$\pm$18.79  & 57.38$\pm$17.55 & 77.19$\pm$21.69	
            & 7.09$\pm$12.83   & 22.35$\pm$34.12
            \\ \hline
            \hline
            basic+dual-multiscale & $\surd$ & $\surd$ & $\times$
            & 71.53$\pm$18.85  & 58.21$\pm$17.55 & 78.29$\pm$21.51	
            & 6.04$\pm$10.49   & 19.89$\pm$27.70
            \\ \hline \hline
            semi-basic & $\times$ & $\times$ & $\surd$
            & 69.89$\pm$19.48	& 56.45$\pm$18.67 & 75.84$\pm$23.51	
            & 7.60$\pm$16.81	& 19.92$\pm$33.89
            \\ \hline
            semi-multi-dimensional-scale & $\surd$ & $\times$ & $\surd$
            & 71.20$\pm$18.70	& 57.76$\pm$17.35 & 77.15$\pm$21.29	
            & 6.41$\pm$14.07	& 20.26$\pm$36.86
            \\ \hline \hline
            \textbf{Our method (DM${^2}$T-Net)}	 & $\surd$ & $\surd$ & $\surd$
            & \textbf{72.59$\pm$18.55} & \textbf{59.42$\pm$17.09} & \textbf{80.44$\pm$20.19}
            & \textbf{4.45$\pm$8.06}   & \textbf{16.34$\pm$24.76}
            \\

			\bottomrule[1.5pt]
		\end{tabular}}
	\end{center}
	\vspace{-2.5mm}
\end{table*}

%% file: result-ablation-study.tex
\begin{figure*}[t]
	\centering
    \vspace*{0.5mm}
    \begin{subfigure}{0.115\textwidth}
    	\includegraphics[width=\textwidth]{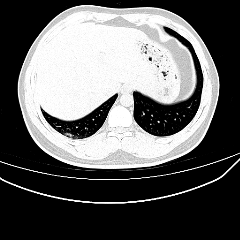}
    \end{subfigure}
    \begin{subfigure}{0.115\textwidth}
    	\includegraphics[width=\textwidth]{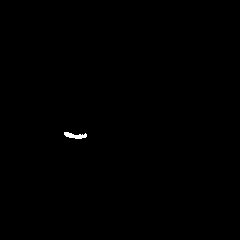}
    \end{subfigure}
    \begin{subfigure}{0.115\textwidth}
    	\includegraphics[width=\textwidth]{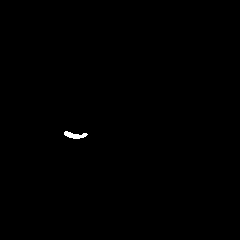}
    \end{subfigure}
    \begin{subfigure}{0.115\textwidth}
    	\includegraphics[width=\textwidth]{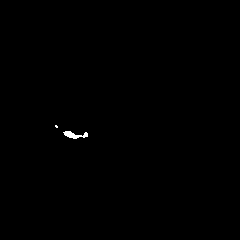}
    \end{subfigure}
    \begin{subfigure}{0.115\textwidth}
    	\includegraphics[width=\textwidth]{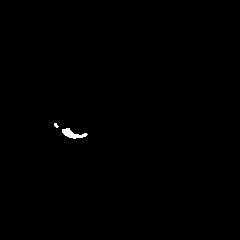}
    \end{subfigure}
    \begin{subfigure}{0.115\textwidth}
    	\includegraphics[width=\textwidth]{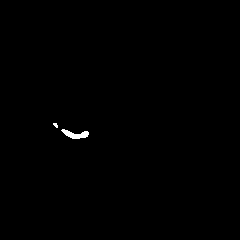}
    \end{subfigure}
    \begin{subfigure}{0.115\textwidth}
    	\includegraphics[width=\textwidth]{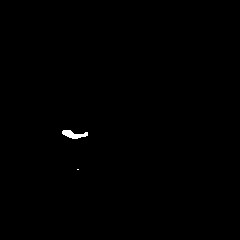}
    \end{subfigure}
    \begin{subfigure}{0.115\textwidth}
    	\includegraphics[width=\textwidth]{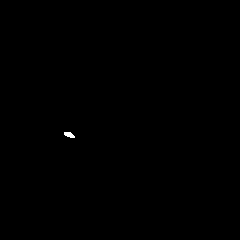}
    \end{subfigure}
    \ \\

	\vspace*{0.5mm}
	\begin{subfigure}{0.115\textwidth}
		\includegraphics[width=\textwidth]{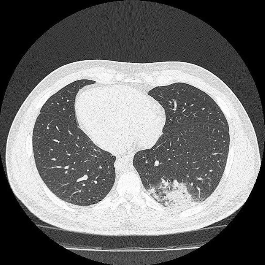}
	\end{subfigure}
	\begin{subfigure}{0.115\textwidth}
		\includegraphics[width=\textwidth]{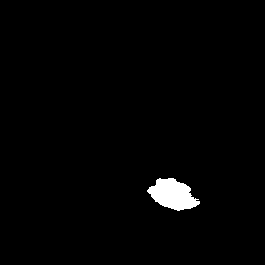}
	\end{subfigure}
	\begin{subfigure}{0.115\textwidth}
		\includegraphics[width=\textwidth]{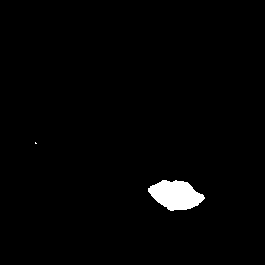}
	\end{subfigure}
	\begin{subfigure}{0.115\textwidth}
		\includegraphics[width=\textwidth]{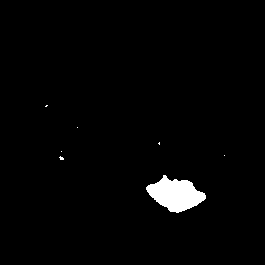}
	\end{subfigure}
	\begin{subfigure}{0.115\textwidth}
		\includegraphics[width=\textwidth]{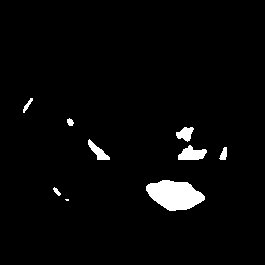}
	\end{subfigure}
	\begin{subfigure}{0.115\textwidth}
		\includegraphics[width=\textwidth]{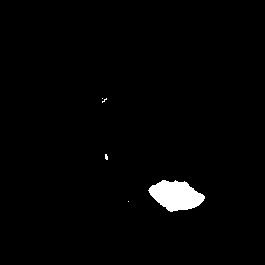}
	\end{subfigure}
	\begin{subfigure}{0.115\textwidth}
		\includegraphics[width=\textwidth]{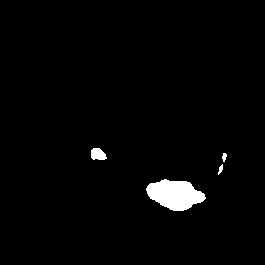}
	\end{subfigure}
    \begin{subfigure}{0.115\textwidth}
		\includegraphics[width=\textwidth]{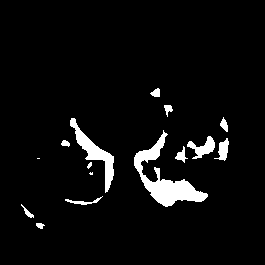}
	\end{subfigure}
	\ \\

    \vspace*{0.5mm}
	\begin{subfigure}{0.115\textwidth}
		\includegraphics[width=\textwidth]{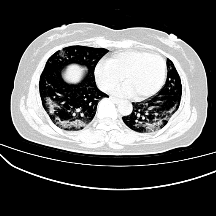}
	\end{subfigure}
	\begin{subfigure}{0.115\textwidth}
		\includegraphics[width=\textwidth]{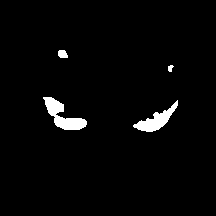}
	\end{subfigure}
	\begin{subfigure}{0.115\textwidth}
		\includegraphics[width=\textwidth]{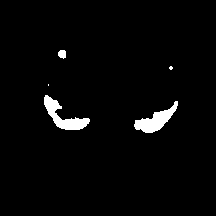}
	\end{subfigure}
	\begin{subfigure}{0.115\textwidth}
		\includegraphics[width=\textwidth]{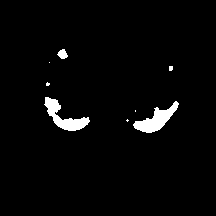}
	\end{subfigure}
	\begin{subfigure}{0.115\textwidth}
		\includegraphics[width=\textwidth]{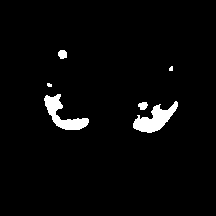}
	\end{subfigure}
	\begin{subfigure}{0.115\textwidth}
		\includegraphics[width=\textwidth]{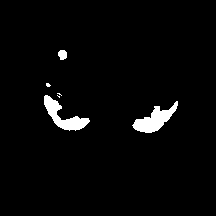}
	\end{subfigure}
	\begin{subfigure}{0.115\textwidth}
		\includegraphics[width=\textwidth]{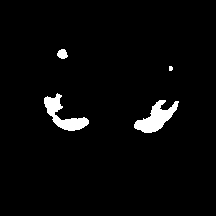}
	\end{subfigure}
    \begin{subfigure}{0.115\textwidth}
		\includegraphics[width=\textwidth]{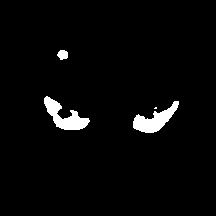}
	\end{subfigure}
	\ \\

     \vspace*{0.5mm}
	\begin{subfigure}{0.115\textwidth}
		\includegraphics[width=\textwidth]{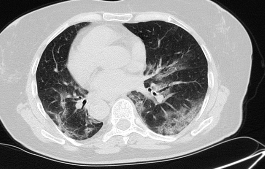}
	\end{subfigure}
	\begin{subfigure}{0.115\textwidth}
		\includegraphics[width=\textwidth]{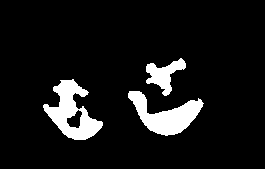}
	\end{subfigure}
	\begin{subfigure}{0.115\textwidth}
		\includegraphics[width=\textwidth]{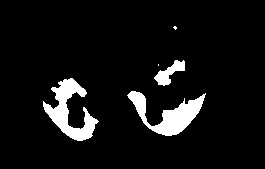}
	\end{subfigure}
	\begin{subfigure}{0.115\textwidth}
		\includegraphics[width=\textwidth]{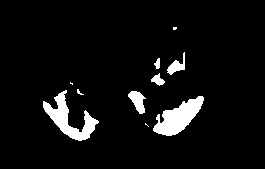}
	\end{subfigure}
	\begin{subfigure}{0.115\textwidth}
		\includegraphics[width=\textwidth]{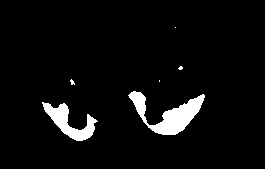}
	\end{subfigure}
	\begin{subfigure}{0.115\textwidth}
		\includegraphics[width=\textwidth]{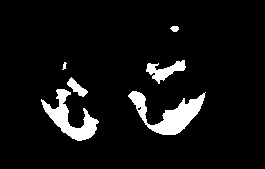}
	\end{subfigure}
	\begin{subfigure}{0.115\textwidth}
		\includegraphics[width=\textwidth]{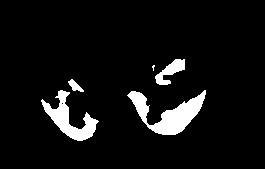}
	\end{subfigure}
    \begin{subfigure}{0.115\textwidth}
		\includegraphics[width=\textwidth]{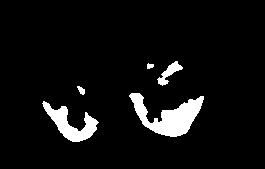}
	\end{subfigure}
	\ \\

    \vspace*{0.5mm}
   \begin{subfigure}{0.115\textwidth}
   	\includegraphics[width=\textwidth]{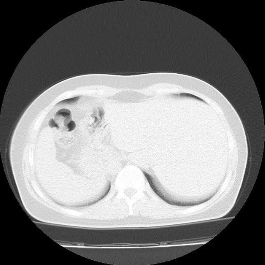} \vspace{-5.5mm} \caption{\footnotesize{inputs}}
   \end{subfigure}
   \begin{subfigure}{0.115\textwidth}
   	\includegraphics[width=\textwidth]{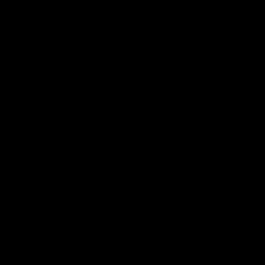}
   	\vspace{-5.5mm} \caption{\footnotesize{GT}}
   \end{subfigure}
   \begin{subfigure}{0.115\textwidth}
   	\includegraphics[width=\textwidth]{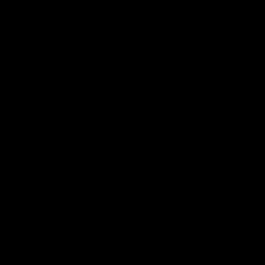} \vspace{-5.5mm}
    \caption{\footnotesize{our method}}
   \end{subfigure}
   \begin{subfigure}{0.115\textwidth}
   	\includegraphics[width=\textwidth]{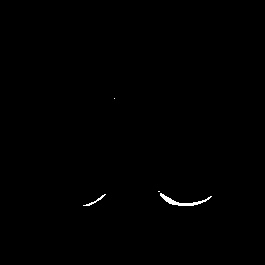} \vspace{-5.5mm}
   \caption{\footnotesize{semi-mdsca}}
   \end{subfigure}
   \begin{subfigure}{0.115\textwidth}
   	\includegraphics[width=\textwidth]{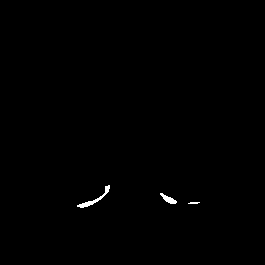}
   	\vspace{-5.5mm} \caption{\footnotesize{semi-basic}}
   \end{subfigure}
   \begin{subfigure}{0.115\textwidth}
   	\includegraphics[width=\textwidth]{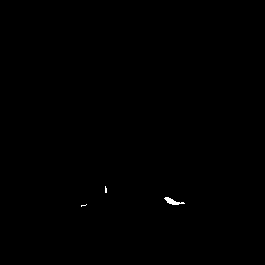}
    \vspace{-5.5mm} \caption{\footnotesize{dual-mulsca}}
   \end{subfigure}
   \begin{subfigure}{0.115\textwidth}
   	\includegraphics[width=\textwidth]{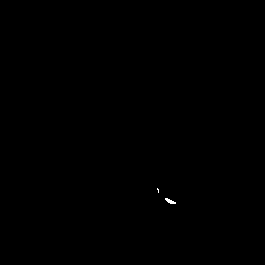}
    \vspace{-5.5mm} \caption{\footnotesize{mul-dimsca}}
   \end{subfigure}
   \begin{subfigure}{0.115\textwidth}
   	\includegraphics[width=\textwidth]{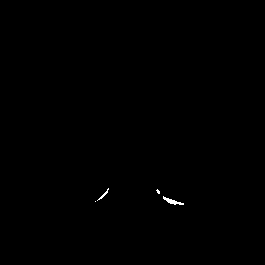}
    \vspace{-5.5mm} \caption{\footnotesize{basic}}
   \end{subfigure}

	\caption{Visual comparisons on COVID-19 infection segmentation results produced by different networks of the ablation study experiment \wjcfinal{on the COVID-19-P20 dataset}. Note that ``mul-dimsca'' ``dual-mulsca'', and ``semi-mdsca'' denote ``basic+multi-dimensional-scale'', ``basic+dual-multiscale'', and ``semi-multi-dimensional-scale'', respectively. }
	\label{fig:ablation_study}
	
\end{figure*}

%% file: table-generalization.tex
\begin{table}[t]
    \centering
    \vspace{-1mm}
	\begin{center}
    \resizebox{1.0\linewidth}{!}{%
		\begin{tabular}{c|c|c|c|c}
			\toprule
			&\multicolumn{2}{c}{$\mathrm{C} \rightarrow \mathrm{M}$}
			&\multicolumn{2}{c}{$\mathrm{M} \rightarrow \mathrm{C}$}\\
			 & Dice $\uparrow$ & HD95  $\downarrow$
			 & Dice $\uparrow$ & HD95  $\downarrow$  \\
			\midrule
			U-Net++~(\cite{zhou2018unet++})
			& 11.09$\pm$11.48 & 51.24$\pm$24.78
			& 23.61$\pm$17.31 & 66.82$\pm$26.13
			\\
			UA-MT~(\cite{yu2019uncertainty})  
			& 43.95$\pm$24.69& 44.95$\pm$34.12
			& 50.24$\pm$17.70 & 51.98$\pm$38.58
			\\
			\midrule
            Ours
            & \textbf{48.81$\pm$23.51} & \textbf{35.34$\pm$30.57}
            & \textbf{51.36$\pm$17.80} & \textbf{41.55$\pm$32.19}\\
            \bottomrule
         \end{tabular}
         }
    \end{center}
    \caption{Generalization analysis, where we use one dataset to train the network and evaluate it on the other one dataset. Let $\mathrm{C}$ and $\mathrm{M}$ denote the COVID-19-P20 and MosMedData datasets, so that $\mathrm{C} \rightarrow \mathrm{M}$ means training the network on $\mathrm{C}$ and evaluating it on $\mathrm{M}$.}
    \label{tab:gl}
\end{table}

%% file: conclusion.tex
\section{Conclusion}
\label{sec::conclusion}
This work has presented a novel COVID-19 lung infection segmentation network from 3D CT volumes by developing a dual multi-scale mean teacher network (DM${^2}$T-Net).
Our key idea is to first develop a multiple dimensional-attention convolutional neural network (MDA-CNN)
to explore multiple dimensional-scale details of the input 3D CT scan.
Moreover, we also employ a semi-supervised system to leverage additional unlabeled data and dual multi-scale information for further boosting 
the COVID-19 infected lung region segmentation. 
%
\wjcfinal{Two datasets for COVID-19 segmentation are collected to evaluate the effectiveness, where our model finally achieves the Dice score of 72.59\% on the COVID-19-P20 dataset and achieved the Dice score of 60.19\% on the MosMedData dataset. The results show that our DM${^2}$T-Net performs better than the state-of-the-art methods by a large margin.}